\documentclass[onecolumn, a4paper,journal,10pt]{IEEEtran}
\usepackage{amsfonts}
\usepackage{graphicx, epsfig,amsmath,amssymb,latexsym,graphics,float}
\usepackage{setspace,subfig}
\usepackage{citesort}
\usepackage{float}
\usepackage[ruled,vlined,linesnumbered]{algorithm2e}
\include{IEEEtran.bib}

\begin{document}

\title{A Sequential Scheme for Large Scale Bayesian Multiple Testing}
\author{Bin~Liu$^{\ast}$, Giuseppe Vinci, Adam C. Snyder, Robert E. Kass
\thanks{B. Liu is with School of Computer Science, Nanjing University of Posts and Telecommunications, Nanjing, China, 210023. Giuseppe Vinci is with Department of Statistics, Carnegie Mellon University, Pittsburgh, U.S., 15213. Adam C. Snyder is with ECE Department, Carnegie Mellon University, Pittsburgh, U.S., 15213. Robert E. Kass is with Department of Statistics, Machine Learning Department, Center for the Neural Basis of Cognition, Carnegie Mellon University, Pittsburgh, U.S., 15213

$^{\ast}$ Correspondence author. E-mail: bins@ieee.org}% <-this % stops a space
%\thanks{Manuscript received XX. X, 2016; revised XX X, 2016.}
}
\maketitle

\begin{abstract}
The problem of large scale multiple testing arises in many contexts, including testing for pairwise interaction among large numbers of neurons. With advances in technologies, it has become common to record from hundreds of neurons simultaneously, and this number is growing quickly, so that the number of pairwise tests can be very large. It is important to control the rate at which false positives occur. In addition, there is sometimes information that affects the probability of a positive result for any given pair. In the case of neurons, they are more likely to have correlated activity when they are close together, and when they respond similarly to various stimuli. Recently a method was developed to control false positives when covariate information, such as distances between pairs of neurons, is available. This method, however, relies on computationally-intensive Markov Chain Monte Carlo (MCMC). Here we develop an alternative, based on Sequential Monte Carlo, which scales well with the size of the dataset. This scheme considers data items sequentially, with relevant probabilities being updated at each step. Simulation experiments demonstrate that the proposed algorithm delivers results as accurately as the previous MCMC method with only a single pass through the data. We illustrate the method by using it to analyze neural recordings from extrastriate cortex in a macaque monkey.
\end{abstract}

\IEEEkeywords% begin{keywords}
Bayesian, massive data, multiple hypothesis testing, neuroscience, Sequential Monte Carlo
%\end{keywords}

\section{Introduction}
\subsection{Multiple Testing}
Statistical hypothesis testing is an extremely powerful approach to knowledge discovery from data \cite{bolton2003iterative,liu2015supporting,webb2016multiple,liu2006statistical}.
Classically, a single proposition $h$ is tested and two opposing hypotheses, namely the null hypothesis ($h=0$) and the alternative hypothesis ($h=1$), are examined. It is often desirable to employ this approach to test not just a single proposition but multiple propositions $h_1,\ldots,h_n$ simultaneously, especially in many burgeoning data-intensive research fields such as genomics \cite{ge2003resampling,ignatiadis2016data}, neurophysiology \cite{durante2016bayesian}, electronic commerce \cite{kohavi2009controlled}, machine learning \cite{webb2016multiple}, medical and epidemiological research \cite{armitage2008statistical,bender2001adjusting}, etc.

In the simplest form of multiple testing, each test yields a summary statistic $z_i$ and the goal is to decide which of the $z_i$s are signals $(h_i=1)$ and which are null $(h_i=0)$ while limiting the number of false positives. The most widely applied approach in many fields is to control the false discovery rate (FDR), i.e., the expected proportion of false positives among those null hypotheses that are rejected \cite{benjamini1995controlling}.
It has been shown that the FDR solution is closely related to the Bayesian posterior $p(h_i=0|\mbox{Data})$, and this alternative formulation can offer advantages in some settings \cite{benjamini1995controlling,efron2012large,muller2006fdr}. To improve testing performance, it is sometimes possible to introduce covariate information that can appropriately modify the prior probability of each particular proposition, and the recently developed Bayesian FDR regression (BFDR) provides a formalism and implementation for this \cite{scott2015false}. However, it relies
on a Markov Chain Monte Carlo (MCMC) based iterative sampling procedure for parameter inference. The MCMC mechanism requires a complete scan of the dataset for each iteration. Hence it does not scale well to large datasets.
\subsection{Neural Interaction Detection: A Large Scale Multiple Testing Problem}
There is great interest in examining the correlated activity among neurons recorded from various networks in the brain while an animal performs a task, as this can furnish clues about the functional operation of these networks
\cite{smith2013spatial,saalmann2011cognitive,smith2008spatial,averbeck2006neural,cohen2011measuring,doiron2016mechanics,yatsenko2015improved}.
However, as the number of neurons being recorded simultaneously increases, statistical methods that scale well with dataset size are needed.
In experiments that will take place in the next few years, tens of millions of pairs of neurons will be recorded simultaneously. Determining the ways in which patterns of correlation change with experimental conditions thus poses a large-scale multiple testing problem.
\subsection{Contribution of this paper}
In this paper we first translate the problem of large scale multiple testing into a sequential parameter updating process.

We then design an efficient algorithm to implement the proposed sequential scheme based on a specific parametric model that incorporates covariates information in modeling the test statistics. This algorithm features novel particle rejuvenation operations designed for updating particles defined on a joint parameter space with both discrete and continuous parameters. Each particle corresponds to a specific hypothesis on the model parameter.
We conduct simulated experiment to show proof-of-concept. Results demonstrate the desirable estimation and testing performance of our algorithm and its scalability to large dataset.

Finally, we apply the proposed algorithm to analyze neural recordings from extrastriate visual cortex in a macaque monkey. The pattern of neural functional connectivity is inferred from the data. The network structure is found to vary with the experimental condition. It is also shown that the sequential feature of our algorithm allows the analyst to inspect this variation in real time.
The results we provide are illustrative, as we anticipate datasets that are orders of magnitude larger in the near future.
\section{Model}
We adopt a parametric model to describe the relationship between the test statistics and the covariates. This model is designed as follows \cite{scott2015false}
\begin{eqnarray}
z_i & \sim & c(x_i) \cdot f_1(z_i)+\{1-c(x_i)\} \cdot f_0(z_i)\\ %\nonumber
f_0(z)&=& \mathcal{N}(z|\mu_0,\sigma_0^2) \\ %\nonumber
f_1(z)&=& \int_{\mathcal{R}}\mathcal{N}(z|\mu_0+\theta,\sigma_0^2)\pi(\theta)d\theta \\ %\nonumber
c(x_i) &=& \frac{1}{1+\exp\{-s(x_i)\}} \\ %\nonumber
s(x_i) &=& \beta_0+\sum_{j=1}^J\beta_j\cdot x_i^j,
\end{eqnarray}
where $i$ is the index of the observations, $c(\cdot)\in(0,1)$ denotes the prior probability that $z_i$ is a signal, $x_i$ denotes the covariates,
which affects $c$ in a way as specified by Eqn.(4), $J$ is the dimensionality of $x$, $\beta=[\beta_0,\beta_1,\ldots,\beta_J]$ is the regression coefficient,
$f_0$ and $f_1$ respectively denote the
null ($h_i=0$) and alternative ($h_i=1$) distributions,
and $\pi(\theta)$ is a $K-$component mixture of Gaussians defined to be $\pi(\theta)\triangleq \sum_{j=1}^K w_j\mathcal{N}(\theta|\mu_j,\sigma_j^2)$.
The above model can be rewritten in a hierarchical form
\begin{eqnarray}
(z_i|\theta_i)&\sim & \mathcal{N}(\mu_0+\theta_i,\sigma_0^2)\\
(\theta_i|h_i)&\sim & h_i\cdot\pi(\theta_i)+(1-h_i)\cdot\delta_0\\
\mbox{P}(h_i=1)=c(x_i)&= &\frac{1}{1+\exp\{-(\beta_0+\sum_{j=1}^J\beta_j\cdot x_i^j)\}}
\end{eqnarray}
where $\delta_{a}$ denotes the Dirac-delta function located at $a$.
Such a hierarchical model together with priors for $\beta$ and the unknown distribution
$\pi(\theta)$, defines a joint posterior distribution over all model parameters.
A full Bayesian approach, namely BFDR, has been developed to sample from the posterior \cite{scott2015false}.
As a Bayesian approach, the BFDR method provides a natural way to
quantify uncertainty about the regression function $s(x)$ and $\pi(\theta)$
jointly, while this is counterbalanced by the additional computational
complexity of the fully Bayesian method. As mentioned above in the Introduction section, the main building block of the BFDR method is MCMC, which consists of numerous iterations and each iteration requires the whole dataset to be processed.
The nature of the MCMC methods also determines that, if new data items are added into the dataset, the whole sampling and computation process should be re-started from scratch to update the testing result. This cumbersome computing feature makes the BFDR method unsuitable for large-scale multiple testing problems. To this end, we propose a sequential computation scheme in the next section, which is also full Bayesian, while scaling well to large dataset.
\section{The Proposed Sequential Scheme for Multiple Testing}\label{sec:online}
Given a parametric model of the test statistics, here we show how to perform multiple testing in a sequential manner. Denote $\phi$ as the unknown model parameter that must be estimated. Denote $t$, $t\in\mathbb{Z}, t>0$, as a time variable and let $\eta_t\triangleq p(\phi|z_1,\ldots,z_t)$
represent the target probabilistic density function (pdf) at time step $t$. As the proposed scheme processes data items one by one in a sequential manner, $z_t$ essentially represents the $t$-th data item that is processed. If the sample size of the dataset to be analyzed is $n$, then the target pdf of our final concern is $\eta_n$.

We now consider a series of target densities, $\eta_1,\ldots,\eta_n$, and focus on the question how to sample from them sequentially.
Suppose that, at time $t+1$, a sampling mechanism is not readily available for the target density $\eta_{t+1}$, but one is available for another
sampling density $\eta_t$, then we can use importance sampling and write \cite{chopin2002sequential}
\begin{eqnarray}\label{eqn:is}
E_{\eta_{t+1}}\{q(\phi)\}&=&\int_{\Phi}q(\phi)\eta_{t+1}(\phi)d\phi\\
&=&\int_{\Phi}q(\phi)\frac{\eta_{t+1}(\phi)}{\eta_t(\phi)}\eta_t(\phi)d\phi\\
&=&\lim_{M\rightarrow\infty}\frac{\sum_{m=1}^M q(\phi_m)\hat{\omega}_m}{\sum_{m=1}^M\hat{\omega}_m}
\end{eqnarray}
where $q$ is any measurable function such that the left hand side of Eqn.(\ref{eqn:is}) exists,
$\phi_m$ is a draw from $\eta_t(\phi)$
and $\hat{\omega}_m\propto\eta_{t+1}(\phi_m)/\eta_t(\phi_m)$.
The weights $\hat{\omega}_m$ can be known only up to a multiplicative constant, which is
cancelled by the denominator $\sum_{m=1}^M\hat{\omega}_m$. According to importance sampling theory,
we have
\begin{equation}
M^{\frac{1}{2}}\left\{\sum_{m=1}^M\omega_m q(\phi_m)-E_{\eta_{t+1}}\{q(\phi)\}\right\}\rightarrow \mathcal{N}(0,V(q))
\end{equation}
in distribution, where $V(q)=\mbox{var}_{\eta_t}[\{\eta_{t+1}(\phi)/\eta_t(\phi)\}\{q(\phi)-E_{\eta_{t+1}}q(\phi)\}]$ \cite{chopin2002sequential},
$\omega_m=\hat{\omega}_m/\sum_{i=1}^M\hat{\omega}_i$.
The weighting operation thus shifts the target density of the particle set from $\eta_t$ to $\eta_{t+1}$.
Given an initial target density $\eta_0$, we perform the above sampling and weighting operations sequentially, shifting the target of interest
from $\eta_0$ to $\eta_1$, from $\eta_1$ to $\eta_2$ and
so on, until from $\eta_{n-1}$ to $\eta_n$. Such a sequential importance sampling (SIS) procedure constitutes the basic building block of a
Sequential Monte Carlo (SMC) method, which has gained a lot of attentions in the fields of statistics and signal processing
\cite{del2006sequential,arulampalam2002tutorial}.

In the basic SIS procedure, as long as the target density evolves one time step forward, say from $t$ to $t+1$,
a reweighting operation $\omega_m'=\hat{\omega}_m\eta_{t+1}(\phi_m)/\eta_t(\phi_m)$ will be
performed. As this reweighting operation iterates, fewer and fewer particles will retain significant
weights. This phenomenon is termed particle degeneracy or impoverishment. A common strategy to avoid that is to perform
resampling followed by a rejuvenation step \cite{gilks2001following}. The resamping operation duplicates particles with higher weights and discards
those with lower weights \cite{Li2015Resampling,Hol2006on}. The rejuvenation step is operated on the resampled particles, moving them according to a
Markov transitional kernel with stationary distribution set to be the target density at current time step. In this way, the target density of the
particles is not changed, but more refreshed particles are generated. Therefore the phenomenon of particle degeneracy may be reduced a lot. Under mild regularity conditions, rigorous convergence results for general SMC methods have been proved \cite{crisan2002survey}.

Here we propose to employ SMC, instead of MCMC, to perform multiple testing. One iteration of the proposed scheme is presented in Algorithm 1 as follows.
\begin{algorithm}[!htb]
Weighting: update the the particle weights by $\hat{\omega}_m=\eta_{t+1}(\phi_m)/\eta_t(\phi_m), \omega_m=\hat{\omega}_m/\sum_{j=1}^M\hat{\omega}_j$, for $m=1,\ldots,M$\;
Resampling: Sample $\phi_m^r\sim\sum_{j=1}^M\omega_j\delta_{\phi_j}$, for all $m$, $1\leq m\leq M$.
Set $\phi_m=\phi_m^r$ and $\omega_m=1/M$
for $m=1,\ldots,M$\;
Particle rejuvenation: draw $\phi_m'\sim T_{t+1}(\phi_m,\cdot)$ for $m=1,\ldots,M$,
 where $T_{t+1}$ is a transition kernel with stationary distribution $\eta_{t+1}$\;
Loop: Set $t=t+1$. If $t<n$, set $\phi_m=\phi_m'$, for $m=1,\ldots,M$, and return to Step 1; otherwise, go to Step 5\;
Testing: Set $\hat{\phi}=\phi_m$, $m=\underset{1\leq j\leq M}{\arg\max} (\omega_j)$. Calculate the posterior probability that $z_i$ is a signal $p(h_i=1|\hat{\phi})$ for all $i$, $1\leq i\leq n$. Declare those $z_i$s corresponding with $p(h_i=1|\hat{\phi})>0.5$ as signals (i.e., the alternative hypotheses) and the others as the null hypotheses.
\caption{\label{algo:SMC}The Proposed Sequential Scheme for Multiple Testing (at time step $t+1, t\geq0$)}
\end{algorithm}
\section{Algorithm Design}
Here we design an efficient algorithm to implement the proposed scheme presented in Section 3 based on the model described in Section 2.
Recall that the model is defined by Eqns. (1)-(5). The unknown parameter to be estimated is
\begin{equation}
\phi=\{\beta, K, \mu_0, \sigma_0, \mu_1, \ldots, \mu_K, \sigma_1, \ldots, \sigma_K, w_1, \ldots, w_{K-1}\}.
\end{equation}
As $w_K=1-\sum_{j=1}^{K-1}w_j$, it is not included in $\phi$.
In contrast with standard Bayesian sequential filtering problems, one notable feature of our problem lies in that the dimensionality of the model parameter is a variable, whose value depends on the number of mixing components $K$.
Therefore, to implement the sequential scheme presented in Algorithm 1, we should take into account of the cases, wherein the particles are assigned with different $K$ values and thus have different dimensions. To our knowledge, these is seldom algorithm that can operate on a set of particles with different dimensions in the SMC literature. Additionally, when $K$ is a large number, the dimensionality of $\phi$ becomes large, then a traditional SMC method may collapse due to the curse of dimensionality \cite{bengtsson2008curse}. Hence it is far from trivial to develop an efficient algorithm to implement the proposed scheme based on the model we are concerned here. We describe our algorithm design in what follows.
\subsection{The weighting and resampling steps}
The weighting step involves the calculation of $\omega_m$. Based on Bayesian theorem, we have $\eta_{t+1}(\phi)=\eta_{t}(\phi)\cdot p(z_{t+1}|\phi)/C$, where $C$ is the normalizing constant. Hence we just set $\hat{\omega}_m=p(z_{t+1}|\phi_m)$, which can be calculated directly based on Eqns.(1)-(5), and then we get $\omega_m=\hat{\omega}_m/\sum_{j=1}^M\hat{\omega}_j$. To implement the resampling step, we select the unbiased residual resampling method \cite{liu1998sequential}. Other widely used resampling methods include the multinomial selection method \cite{gordon1993novel} and the stratified resampling method \cite{carpenter1999improved}.
% and the stratified resampling \cite{carpenter1999improved}.
\subsection{Particle rejuvenation step}
We develop an efficient approach to implement the particle rejuvenation step of Algorithm 1. It is the ad hoc design of the particle rejuvenation algorithm that discriminates our SMC sampler from all the other related SMC methods in the literature. First we separate the components of $\phi$ into two parts. Let $\phi=\{\phi^1,\phi^2\}$,
where $\phi^1=\beta$ and $\phi^2$ is composed of the rest of parameters in $\phi$. So $\phi^1$ is of fixed dimensionality and the parameters included
in $\phi^2$ correspond to a mixture model that consists of the null distribution and the alternative distribution.
In the model defined with Eqns. (1)-(5), the null distribution is normal and the alternative distribution is a $K$-component normal mixture. Hence the dimensionality of $\phi^2$ is variable depending on the value of $K$.
We design rejuvenation operations for $\phi^1$ and $\phi^2$, respectively. Suppose that $\phi_m=\{\phi_m^1,\phi_m^2\}$, a random draw from $\eta_t$, is available, we now take this particle as an example to describe the designed particle rejuvenation operations that are performed at time step $t+1$.
\subsubsection{Particle rejuvenation operations for $\phi_m^2$}
Now we describe how to update $\phi_m^2$ based on the new data item $z_{t+1}$ at current time step.
First we calculate the prior probability that $z_{t+1}$ is a signal, through Eqn. (4), based on the current hypothesis on the model parameter $\phi_m$. If this probability is less than 0.5, we allocate $z_{t+1}$ into the null distribution associated with the $m$th particle, represented by $f_{0,m}$, and update its parameters $\mu_{0,m}$ and $\sigma_{0,m}$ as follows
\begin{eqnarray}
\mu_{0,m}&=&(1-\alpha_0)\cdot\mu_{0,m}+\alpha_0\cdot z_{t+1}\\
\sigma_{0,m}&=&\sqrt{(1-\alpha_0)\cdot\sigma_{0,m}^2+\alpha_0\cdot(z_{t+1}-\mu_{0,m})^2},
\end{eqnarray}
where $\alpha_0=1/(1+N_{0,t,m})$ denotes the weight assigned to $z_{t+1}$. $N_{0,t,m}$ denotes the number of $z_i$s that were allocated into $f_{0,m}$ in the previous $t$ time steps.

If the aforementioned prior probability is greater than or equal to 0.5, we allocate $z_{t+1}$ into $f_{1,m}$. Now we present how to update the parameter value of $f_{1,m}$ if $z_{t+1}$ is allocated into it.
We develop an online $K$-means approximation method to update the parameter value of $f_{1,m}$. First the test statistic $z_{t+1}$ is checked against the existing $K_m$ Gaussian distributions of $f_{1,m}$, until a match is found. A match is defined as the value of $z_{t+1}$ falling within 2.5 standard deviations of a distribution. The threshold 2.5 is also used in an online method for background modeling in visual tracking problems \cite{stauffer1999adaptive}.
Then we adjust the prior weights of the $K_m$ distributions, $w_{1,m},\ldots,w_{K,m}$, as follows
\begin{equation}
w_{k,m}=(1-\alpha_1)\cdot w_{k,m}+\alpha_1\cdot M_{k,m}, k=1,\ldots,K_m,
\end{equation}
where $M_{k,m}$ is 1 for the model which matched and 0 for the remaining models, $\alpha_1=1/(1+N_{1,t,m})$ is the weight assigned to $z_{t+1}$. $N_{1,t,m}$ denotes the number of $z_i$s that were allocated into $f_{1,m}$ in the previous $t$ time steps.
If none of the $K_m$ distributions match $z_{t+1}$, a new probable distribution is added to $f_{1,m}$, with $z_{t+1}$ as its mean value,
an initially high variance, and a prior weight set at $\alpha_1$. As a new component is added, $K_m$ is updated with $K_m=K_m+1$.
After the above updating operation, the weights of mixture components are renormalized. The $\mu$ and $\sigma$ parameters of unmatched mixture components
remain the same. The parameters of the component that matches $z_{t+1}$ are updated as follows
\begin{eqnarray}
\mu_{k,m}&=&(1-\varrho)\cdot\mu_{k,m}+\varrho\cdot z_{t+1}\\
\sigma_{k,m}^2&=&(1-\varrho)\cdot\sigma_{k,m}^2+\varrho\cdot(z_{t+1}-\mu_{k,m})^2,
\end{eqnarray}
where
\begin{equation}
\varrho=\frac{\alpha_1}{\alpha_1+w_{k,m}}.
\end{equation}

The above design of the mixture updating operations is inspired by a $K$-means approximation method used for background modeling in visual object tracking problems \cite{stauffer1999adaptive}, while our method is different from that in \cite{stauffer1999adaptive} in several
ways. First, our method allows the number $K$ to be adaptable, while, the value of $K$ is fixed in \cite{stauffer1999adaptive}. Second,
in our method, the learning rates, $\alpha_0$, $\alpha_1$ and $\varrho$, are controlled through $N_{0,t,m}$ and $N_{1,t,m}$ adaptively. In contrast, for the $K$-means approximation method in \cite{stauffer1999adaptive}, these learning rates are set as free parameters, whose values are specified empirically without clear explanations.
Our method uses $N_{0,t,m}$ and $N_{1,t,m}$ to record the numbers of $z_i$s that have been involved in the previous $t$ time steps for constructing $f_{0,m}$ and $f_{1,m}$, respectively. When $z_{t+1}$ arrives, the amount of the relative contribution of $z_{t+1}$ in updating $f_{0,m}$ (resp. $f_{1,m}$) is encoded by $\alpha_0$ (resp. $\alpha_1$), which is a function of $N_{0,t,m}$ (resp. $N_{1,t,m}$). As the values of $N_{0,t,m}$ and $N_{1,t,m}$ increase over time, the relative contribution of $z_{t+1}$ in updating the parameters of $f_{0,m}$ and $f_{1,m}$ is weakened automatically over time.
\subsubsection{Particle rejuvenation operations for $\phi_m^1$}
After updating $\phi_m^2$, we rejuvenate $\phi_m^1$, i.e., $\beta_m$.
First, based on the set of resampled $\beta_i$s, $i=1,\ldots,M$, a kernel smoothing method \cite{balakrishnan2006one} is adopted here to approximate the posterior $p(\beta|z_1,\ldots,z_{t+1})$ by a series of Gaussian kernel functions:
\begin{equation}\label{fun:kernel}
p(\beta|z_1,\ldots,z_{t+1})\simeq\sum_{i=1}^M\mathcal{N}(\beta|\hat{\beta}_i,b^2Q)
\end{equation}
where $Q$ is the sample Monte Carlo variance, $b=\left(\frac{4}{(d+2)M}\right)^{\frac{1}{d+4}}$ is the kernel
bandwidth, $d$ is the dimensionality of $\beta$, $\hat{\beta}_i=a\cdot\beta_i+(1-a)\cdot\bar{\beta}$,
$a=\sqrt{(1-b^2)}$ and $\bar{\beta}$ is the current Monte Carlo mean $\beta$ value.  This choice of bandwidth has been proved to be asymptotically
optimal for multivariate-normal distribution cases \cite{stavropoulos2001improved}.
Now we set $\beta_m$ as a random draw from $\mathcal{N}(\beta|\hat{\beta}_m,b^2V)$. All particle weights of these rejuvenated $\beta_m$s are equally set at $1/M$ because the resample step preceded this rejuvenation step. These rejuvenated $\beta_m$s now represent a more diverse set of $\beta$ values, and, under mild conditions, these samples converge to those drawn directly from the target density with theoretic guarantees \cite{silverman1986density,balakrishnan2006one}.
\subsection{Initialization}\label{sec:initialization}
To run Algorithm 1, $\eta_0(\phi)$ along with a weighted set of particles $\{\phi_m,\omega_m\}_{m=1}^M$ drawn from it, $N_{0,0,m}$ and $N_{1,0,m}$ must be initialized.
We consider two cases for initialization. In the first case, we assume that a historical dataset, which is collected under the same condition as that of the current dataset, has been processed and a corresponding posterior is available, then we just set $\eta_0(\phi)$, $N_{0,0,m}$ and $N_{1,0,m}$ based on this posterior. In another word, the posterior corresponding to the historical dataset is taken as the prior for analyzing the current dataset.
In the other case, there is no information from a historical dataset available for use. Then we resort to domain knowledge to do initialization. If the domain knowledge contains little information, we can select a noninformative prior, e.g., a diffusing prior over the parameter space, for use. If there is rich domain knowledge, then prior elicitation techniques can be used for distilling the prior from the domain knowledge \cite{albert2012combining,dey2012practical}.
Specifying the prior is an inevitable step to apply Bayesian methods, while, it is noteworthy to mention that, as more and more observations have been processed, the impact of the prior on the final posterior inference will become weaker and weaker, as long as the prior is properly formulated \cite{gelman2014bayesian}.
%Therefore, for cases with massive dataset concerned here, the impact of prior on the final testing result can be relatively small as compared with that for cases with sparse data.
\section{Simulated Data}\label{sec:simu}
The proposed SMC sampler algorithm is first tested on synthetic dataset in order to
verify its efficiency and merits, and confirm its ability to
yield meaningful results when applied to experimental spike train data. A number $n=10000$ of test statistics $z_i, i=1,\ldots,n$, are drawn according to the covariate-dependent
mixture model defined with Eqns.(1)-(5). The dataset has two covariates $x=(x^1,x^2)$ and
the regression coefficient $\beta=(-3.5, \frac{\sqrt{2}}{2}, \frac{\sqrt{2}}{2})$.
The null and the alternative distributions are specified as $f_0(z)=\mathcal{N}(z|0,1)$ and $f_1(z)=\mathcal{N}(z|3,0.5^2)$, respectively.

We select the BFDR method as a baseline for performance comparison, as it represents the
state-of-the-art multiple testing method for the neural interaction detection problem. We implement the BFDR method
in full accordance with its companion R package FDRreg \cite{scott2015false}, where the number of mixing components $K$ in $f_1$ is chosen based on the Akaike information criterion via a preliminary run of an Expectation-Maximization algorithm; the other components of $\phi$ are estimated
by MCMC. The MCMC procedure consists of 2200 iterations of sampling, and the first 200 iterations are taken as the burn-in period.
For the proposed SMC sampler, its parameter setting is presented in Table 1. The parameter value, including the ratio of $N_{0,0,m}$ over $N_{1,0,m}$, is selected according to domain knowledge. The numbers 9 and 1 represent the minimal feasible integers that can be used to initialize $N_{0,0,m}$ and $N_{1,0,m}$.
For both BFDR and SMC, the null distribution is fixed to be a zero-mean Gaussian distribution. For BFDR, the variance of the null distribution is chosen using an offline empirical approach as suggested in \cite{efron2004large}. For both BFDR and SMC, a diffusing uniform prior of $\beta$ is selected for use.
\begin{table}\centering\small
\caption{Parameter setting for the SMC sampler for both the simulated and the real data experiments}
\begin{tabular}{c|c|c|c|c|c|c}
\hline
$N_{0,0,m}$ & $N_{1,0,m}$ & $M$ & $K_m$ & $\mu_{1,m}$ & $\sigma_{1,m}$ & $\sigma_{0,m}$ \\\hline
9 & 1 & $1e4$ & 1 & 3 & $\sqrt{20}$ & 1.5  \\\hline
\end{tabular}
\label{Table:convergence values}
\end{table}

\begin{figure}[t]
\centering
\includegraphics[width=3.6in,height=1.7in]{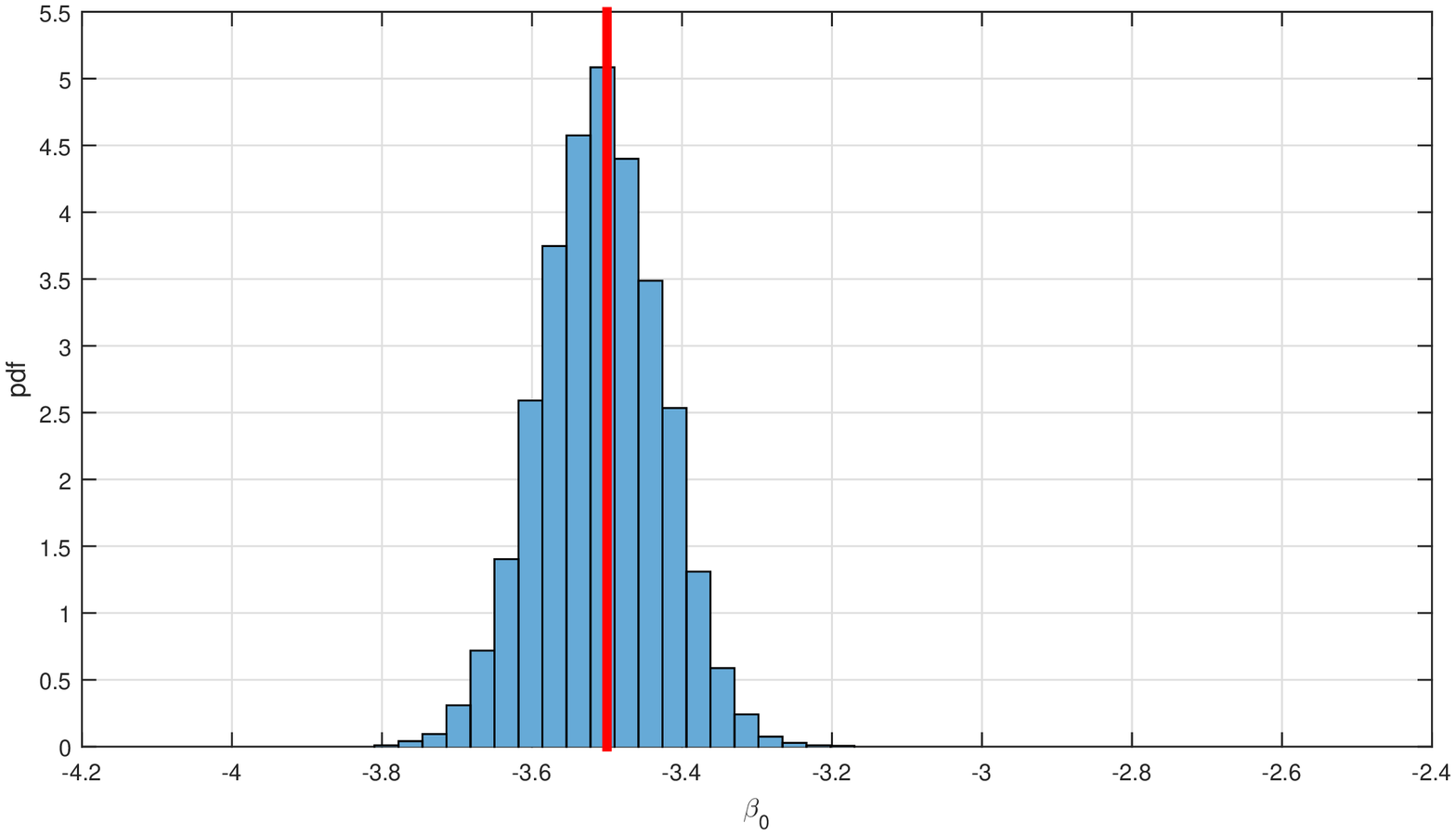}\\
\includegraphics[width=3.6in,height=1.7in]{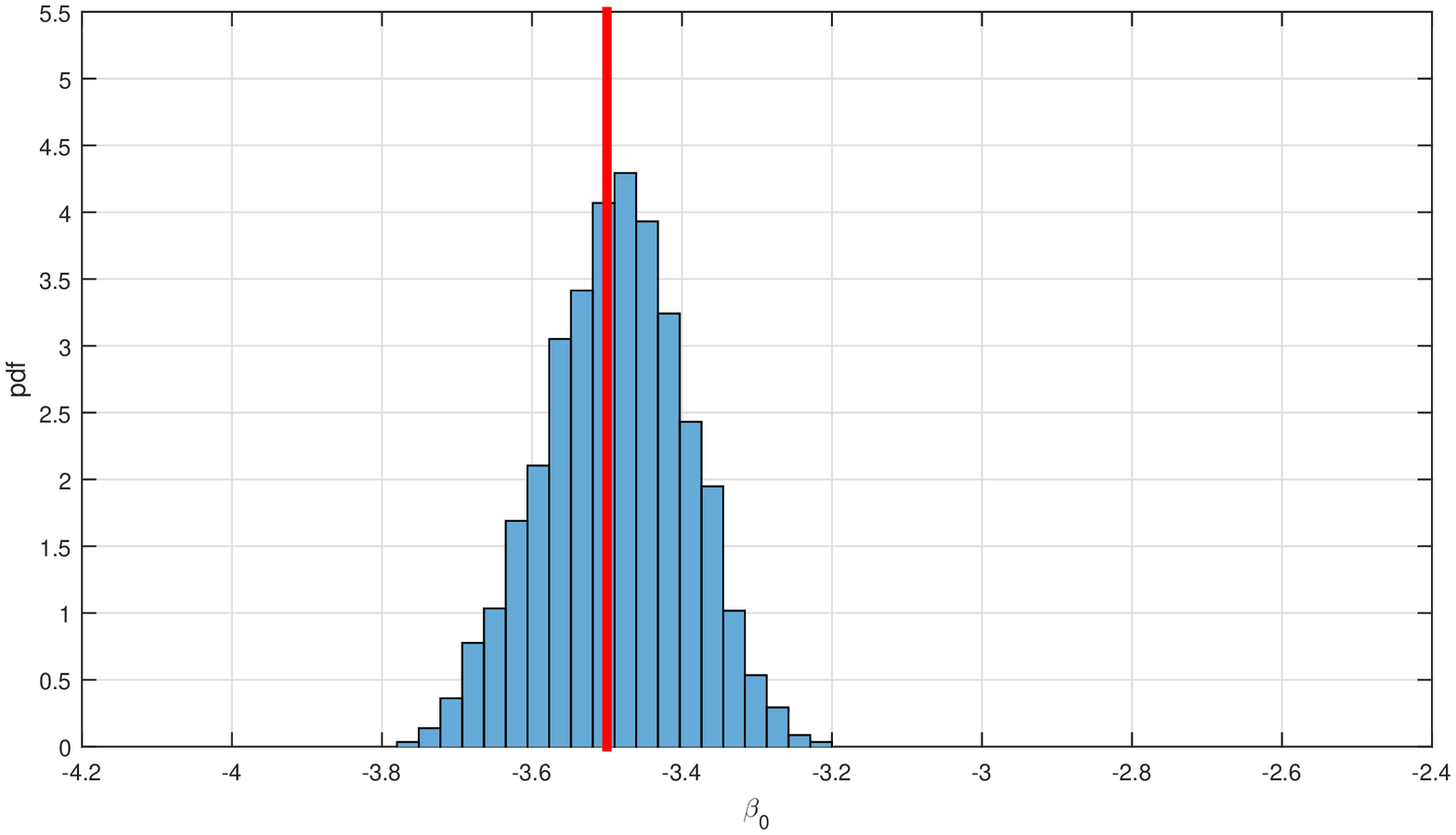}
\caption{Posterior histograms for the estimate of $\beta_0$ given by the SMC sampler (the top panel) and the BFDR method (the bottom panel).
Vertical lines indicate true values for the synthetic data.}\label{fig:hist_beta0}
\end{figure}

\begin{figure}[t]
\centering
\includegraphics[width=2.4in,height=1.75in]{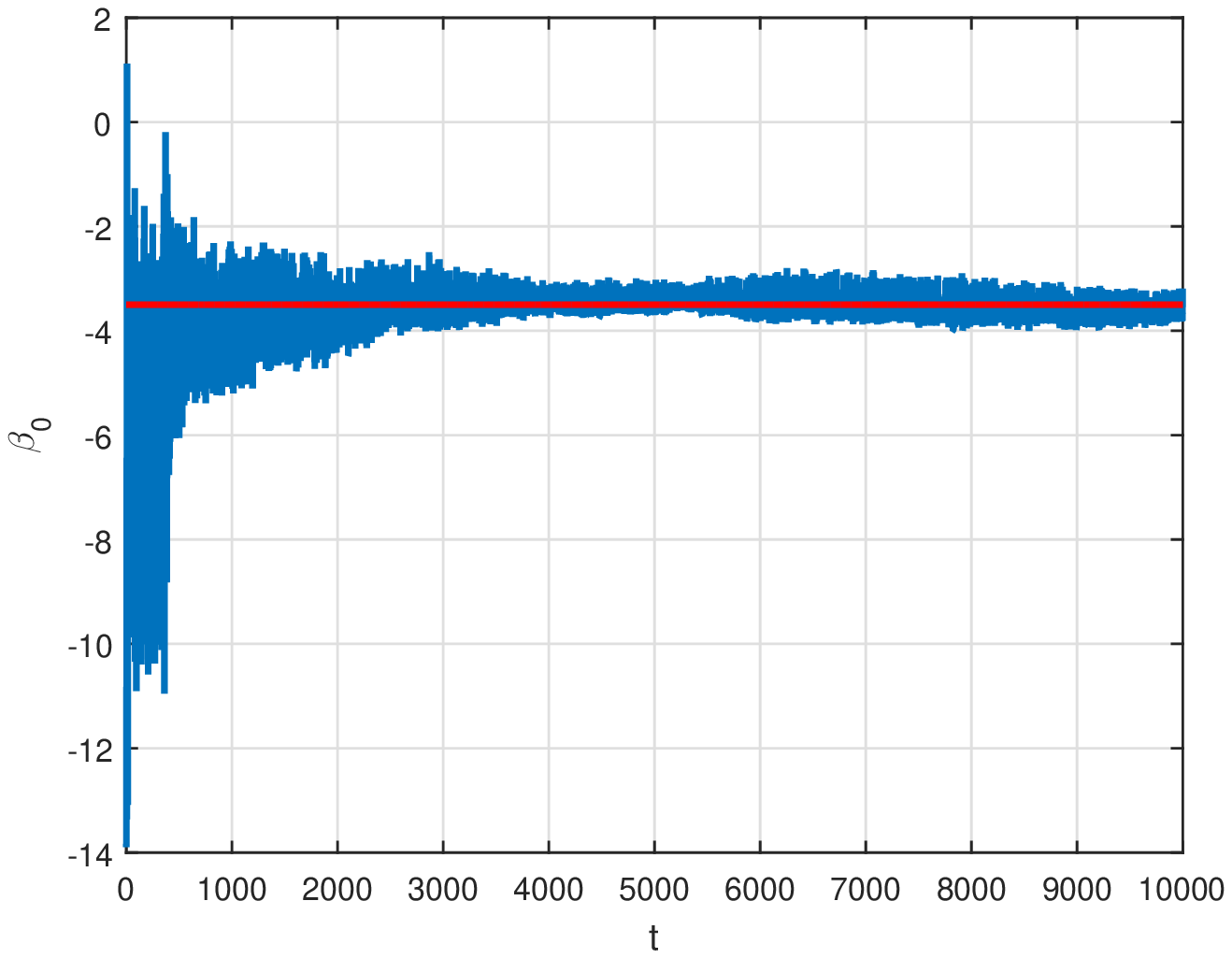}\includegraphics[width=2.4in,height=1.75in]{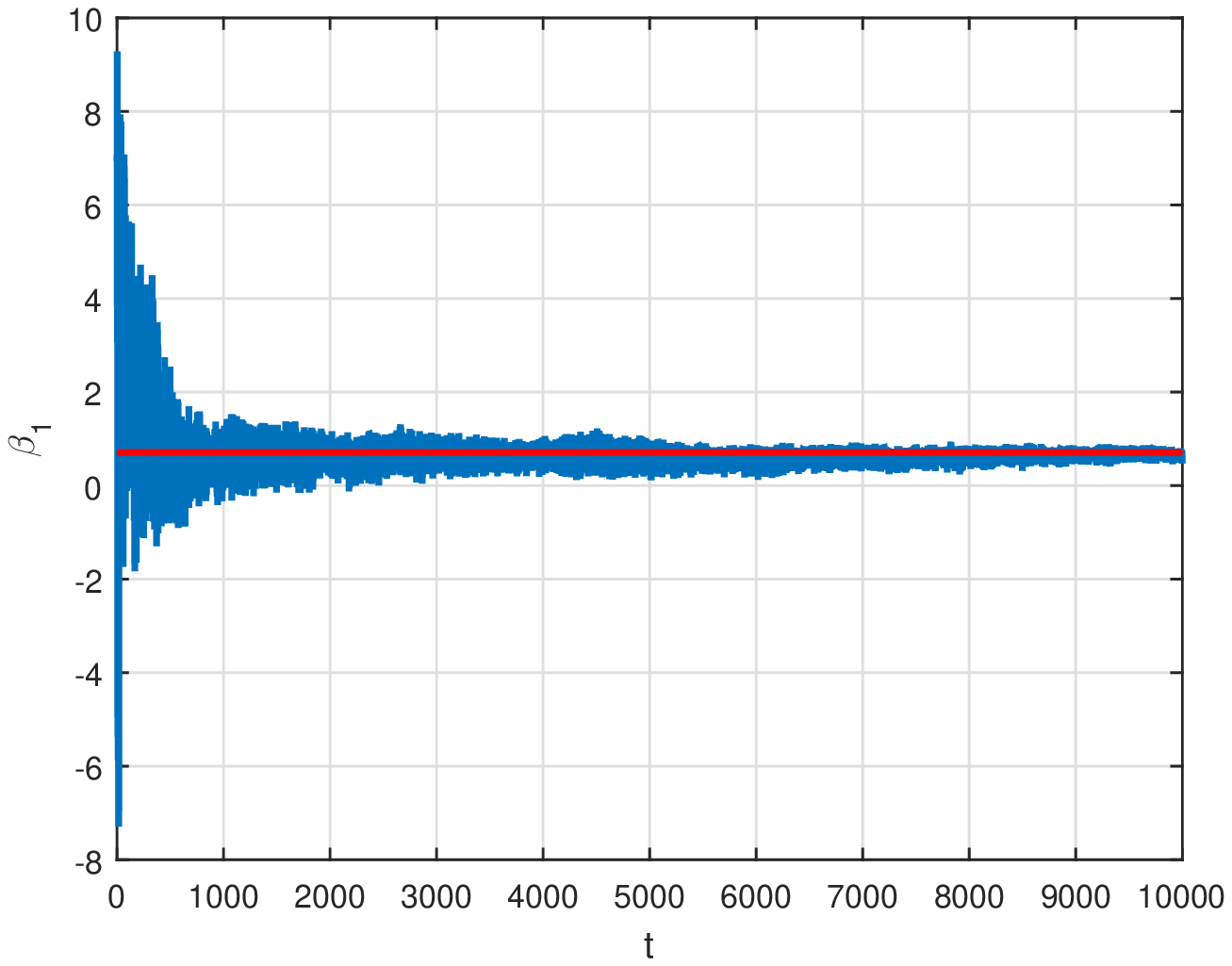}\\
\includegraphics[width=2.4in,height=1.75in]{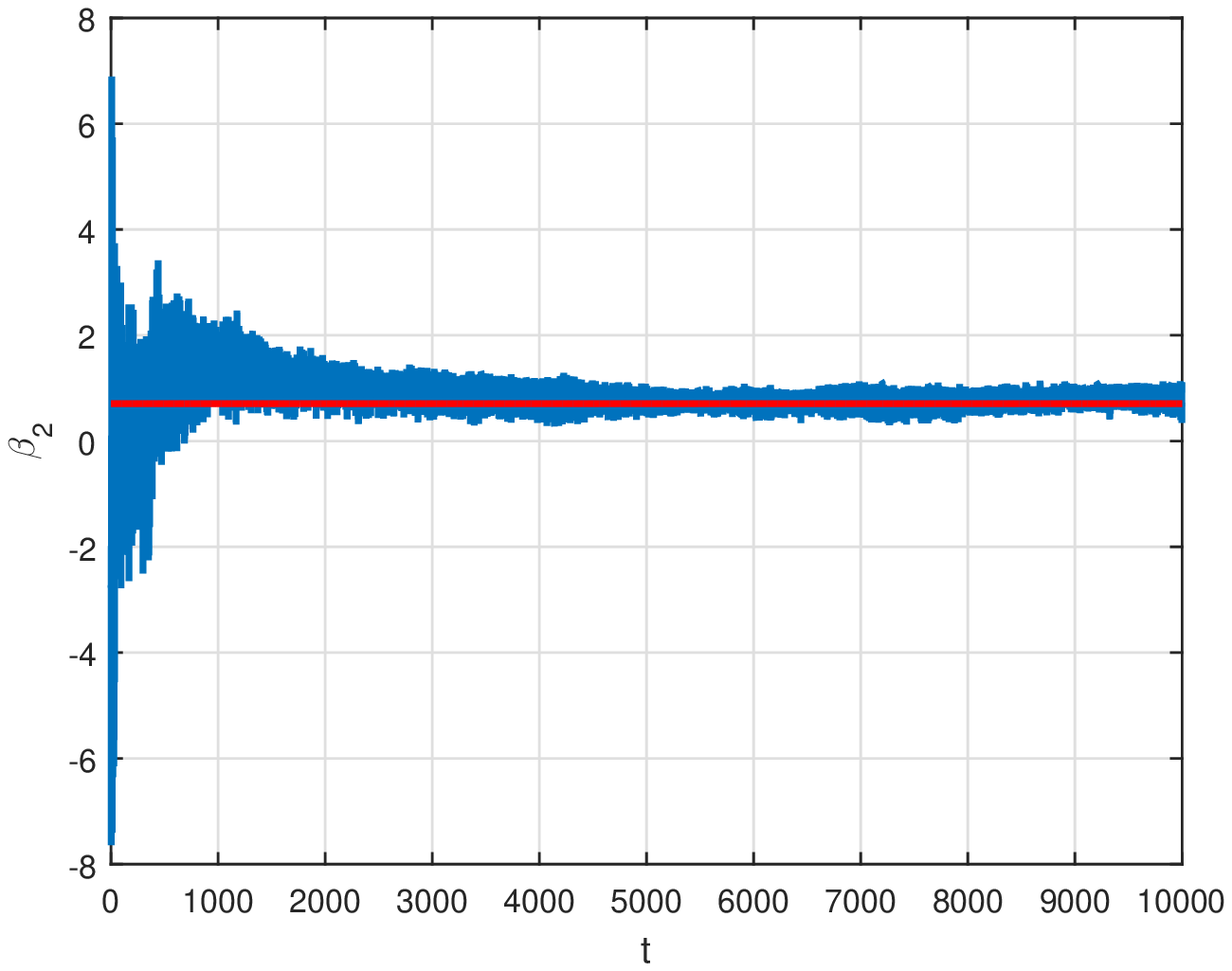}
\caption{Traces of the maximum a posterior estimate for $\beta_0$ (top left panel), $\beta_1$ (top right panel) and $\beta_2$ (bottom panel).
Horizontal lines indicate true values for the synthetic data.}\label{fig:trace_beta}
\end{figure}

The result of the proposed SMC sampler in terms of posterior histograms for the estimate of $\beta_0$
is displayed in Fig.\ref{fig:hist_beta0}. It is shown that, the distributional estimate of $\beta_0$, given by the SMC sampler,
is comparable with that by the BFDR method.
\begin{table}\centering\small
\caption{The estimation result for parameters of the null and the alternative distributions}
\begin{tabular}{c|c|c|c}
\hline
     & $\sigma_0$ & $\mu_1$ & $\sigma_1$ \\\hline
estimated by SMC & 0.9945 & 3.3470 & 0.7930\\\hline
estimated by BFDR & 1.0152 & 3.1070 & 0.1012\\\hline
the true answer & 1 & 3 & 0.5\\\hline
\end{tabular}
\label{Table:estimate_model}
\end{table}
In Fig.\ref{fig:trace_beta}, we plot traces of the maximum a posterior (MAP) estimate of $\beta$ obtained from an example run of the SMC sampler. A good convergence performance of the SMC sampler in estimating $\beta$ is confirmed. The estimation results on $f_0$ and $f_1$ are presented in Table 2. The estimate of $\pi(\theta)$ given by the SMC sampler has two distribution components, while one of them has a tiny weight $0.0027$, so only parameters of the dominant component are listed here. We see that the SMC sampler gives a more accurate estimate on $\sigma_0$ and $\sigma_1$, and a worse estimate on  $\mu_1$, as compared with the BFDR method.

The numbers of detections and errors obtained by the SMC sampler and BFDR method based on the same synthetic dataset are
listed in Tables 3 and 4, respectively. For BFDR, the FDR is controlled roughly at $10\%$ and the real FDR obtained in the experiment is $36/321=11.2\%$. The FDR given by the SMC sampler is $38/331=11.5\%$, which is slightly higher than that of BFDR, while the SMC sampler detects 8 more true alternatives than BFDR. Taking both factors of FDR and detection power into consideration, we argue that the SMC sampler gives almost an equivalent testing performance as that of the BFDR method. The significant difference between the SMC sampler and BFDR lies in that the former accessed each test statistic $z_i$ only once, while the latter accessed each $z_i$ for 2200 times, during this experiment. Assume that a new data item is added to the dataset for analysis. If the BFDR method is under use, the nature of the MCMC sampling mechanism determines that the analysis has to start the whole analysis from scratch, which means that all the historical data have to be re-processed. If the proposed SMC sampler is under use, the analyst can just run one iteration of the SMC sampler, which only needs to access this new data item, without having to retracing any historical data, to update the model parameter as well as the testing result. As demonstrated by the above simulated experiment, the updated result yielded by the SMC sampler will be as accurate as that obtained by applying the BFDR method to analyze the whole updated dataset.
\begin{table}\centering\small
\caption{Number of detections and errors reported by the SMC sampler algorithm}
\begin{tabular}{c|c|c|c}
\hline
     &Declared null &Declared alternative &Total \\\hline
True null & 9539 & 38 & 9577\\\hline
True alternative & 130 & 293 & 423\\\hline
Total & 9669 & 331 & 10000\\\hline
\end{tabular}
\label{Table:error_SMC}
\end{table}
\begin{table}\centering\small
\caption{Number of detections and errors reported by the BFDR method}
\begin{tabular}{c|c|c|c}
\hline
     & Declared null & Declared alternative & Total \\\hline
True null & 9541 & 36 & 9577\\\hline
True alternative & 138 & 285 & 423\\\hline
Total & 9679 & 321 & 10000\\\hline
\end{tabular}
\label{Table:error_BFDR}
\end{table}

\begin{figure}[t]
\centering
\includegraphics[width=3.5in,height=2.3in]{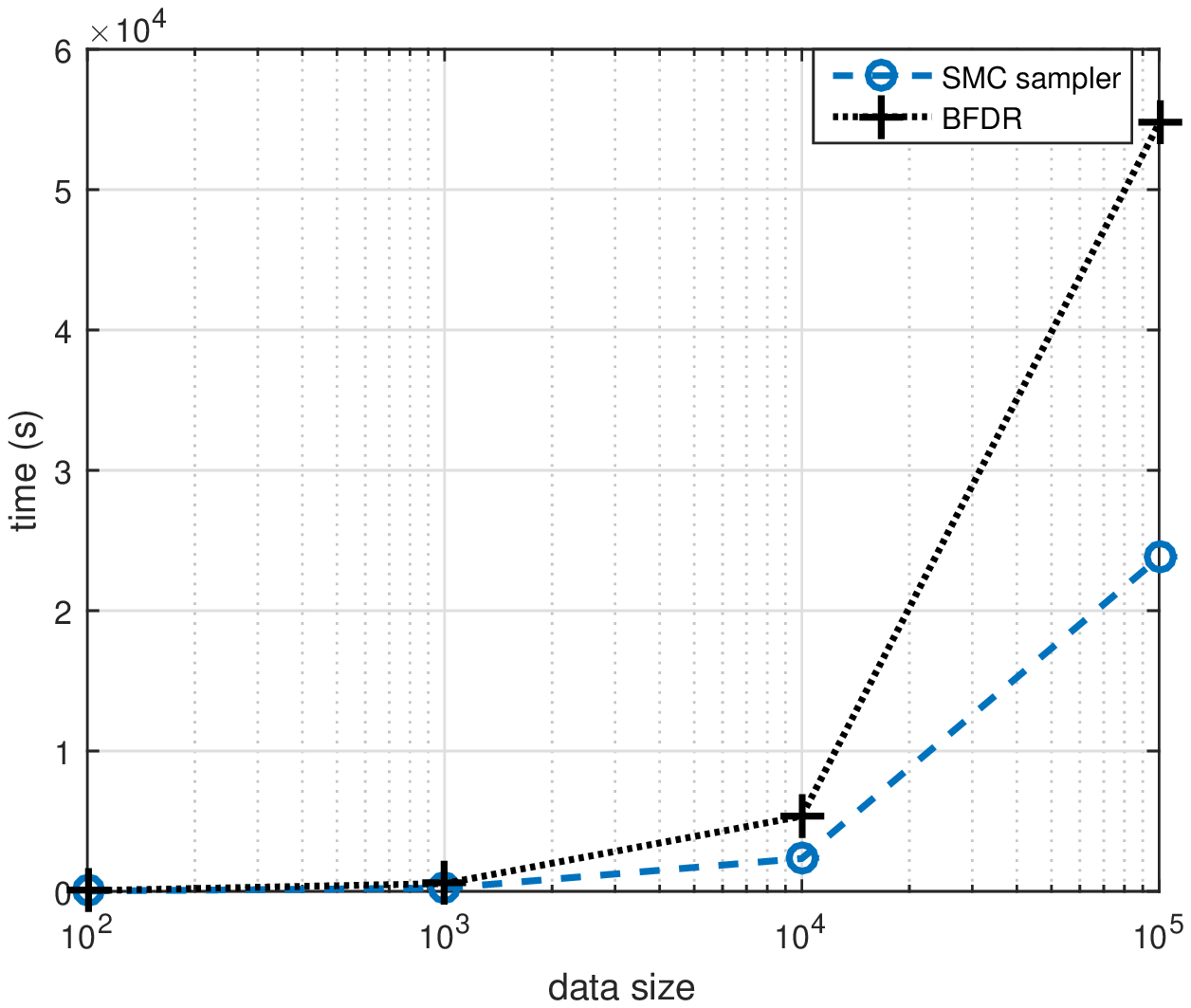}
\caption{Computation time vs. dataset size. }\label{fig:time}
\end{figure}

Finally, we conduct a simulated experiment to compare the proposed SMC sampler and the BFDR method in the computational time. Both methods are coded with MATLAB and run on a 16-core microprocessor. The comparison result is depicted in Fig. \ref{fig:time}. We see that, the greater the amount of data to be analyzed, the more obvious advantages for the SMC sampler in terms of computation time. For the SMC sampler, the particle rejuvenation and weighting
steps are straightforward to parallelise, as they require only independent operations on each particle. Hence these two steps are parallelized in our experiment. The resampling step is more difficult to parallelise, as it requires a collective operation, such as a sum across particle weights. Hence it is not parallelized.
%Recently it has been reported that an alternative resampling scheme can be parallelized by graphic processing unit \cite{murray2016parallel}, which %indicates that the SMC sampler algorithm proposed here can be further accelerated.
The main building block of the BFDR method is an MCMC based iterative sampling procedure, which works in a serial way, therefore there is no straightforward way to parallelise it.
%So the BFDR method is not parallelized in our experiment.
\section{Experimental data}
The experimental data were sampled from a real experimental session where a rhesus macaque monkey was performing a visual spatial selective attention task. Experimental procedures were approved by the Institutional Animal Care and Use Committee of the University of Pittsburgh.
Two task variables are involved. One is orientation. A Gabor stimulus presented in the receptive field is associated with one of two possible orientations, corresponding to 135 degrees and 45 degrees, respectively. We use `ori 1' and `ori 2' to denote these two orientations, respectively. The other task variable is cue. The monkey was cued to either attend towards or away from the receptive field. We use `cue 1' and `cue 2' to denote these two cues, respectively. These two task variables were combined into four stimulus conditions, namely $\{$ori 1, cue 1$\}$, $\{$ori 2, cue 1$\}$, $\{$ori 1, cue 2$\}$ and $\{$ori 2, cue 2$\}$. A number of repeated trials were conducted under each stimulus condition. The numbers of repeated trials are respectively 666, 759, 954, and 1077, for those four stimulus conditions. A Utah array, which consists of a 10 by 10 grid of electrodes, was implanted in area V4 of the monkey's extrastriate visual cortex, and the spike train records were sampled from the same set of neurons across those stimulus conditions (block-randomized).

The test statistics $z_i$s are obtained by applying the Fisher transformation on spike count correlations \cite{kass2014analysis,vinci2016separating,smith2013spatial,smith2008spatial}. Each spike count correlation is a Pearson correlation of spike counts of a pair of neurons. For each test statistic, there are two relevant covariates: (1) inter-neuron distance, measured in micrometers; and (2) tuning-curve correlation (TCC). The dataset to be analyzed consists of four time windows, each corresponding to a specific stimulus condition.

Given the above dataset, the question is how to reveal the neuron network structure, which is determined by neuron interactions, from data. It is also expected to check if this structure varies over the change of the stimulus condition in real time.

\begin{figure}[t]
\centering
\includegraphics[width=4in,height=2.5in]{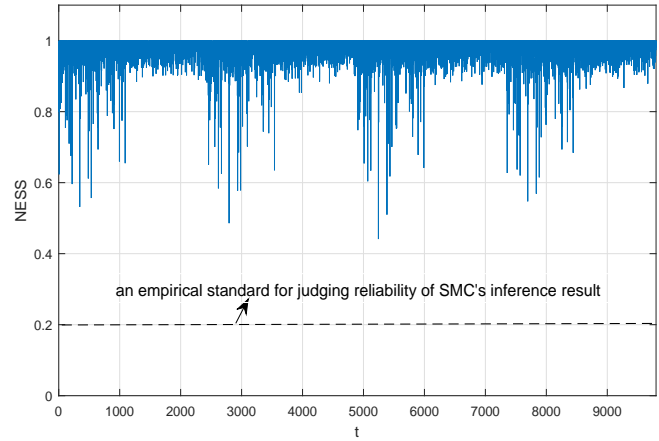}
\caption{Normalized effective sample size of the SMC sampler in processing the experimental data.}\label{fig:ness_real_data}
\end{figure}

The proposed SMC sampler is applied here to test hypotheses on whether pairs of neurons exhibit fine-time-scale ($\sim500ms$) interactions.
We adopt a standard measure termed effective sample size (ESS) to evaluate the reliability of the algorithm's inference result \cite{kong1994sequential}. This measure is defined to be $\mbox{ESS}=1/\{\sum_{m=1}^M(\omega_m)^2\}$, which satisfies $1\leq\mbox{ESS}\leq M$.
The meaning of the ESS can be stated as that, the inference given by the SMC sampler based on the $M$ weighted particles
is equivalent with that obtained based on the number ESS of particles drawn directly from the target distribution \cite{liu1998sequential}. So intuitively, a greater value of ESS indicates a more reliable inference result given by the SMC sampler, and vice versa. Because the ESS is a function of the particle size $M$, we employ the normalized ESS (NESS), $\mbox{NESS}=\mbox{ESS}/M$, in practice. At the transition moments of two neighboring time windows, significant declines in the NESS value appear, because the regularity assumption between the neighboring target pdfs is violated. As long as the NESS value drops below $0.1$, the SMC sampler is re-initialized and then the NESS is re-calculated. The way to initialize SMC sampler here is the same as presented in Section \ref{sec:simu}. We plot the resulting NESS per time step in Fig. \ref{fig:ness_real_data}, which shows that the NESS value maintains at a relatively high level, indicating a reliable inference result of the SMC sampler.
\begin{figure}%[t]
\subfloat[$\{$ori 1,cue 1$\}$]{
\includegraphics[width=2.5in,height=2in]{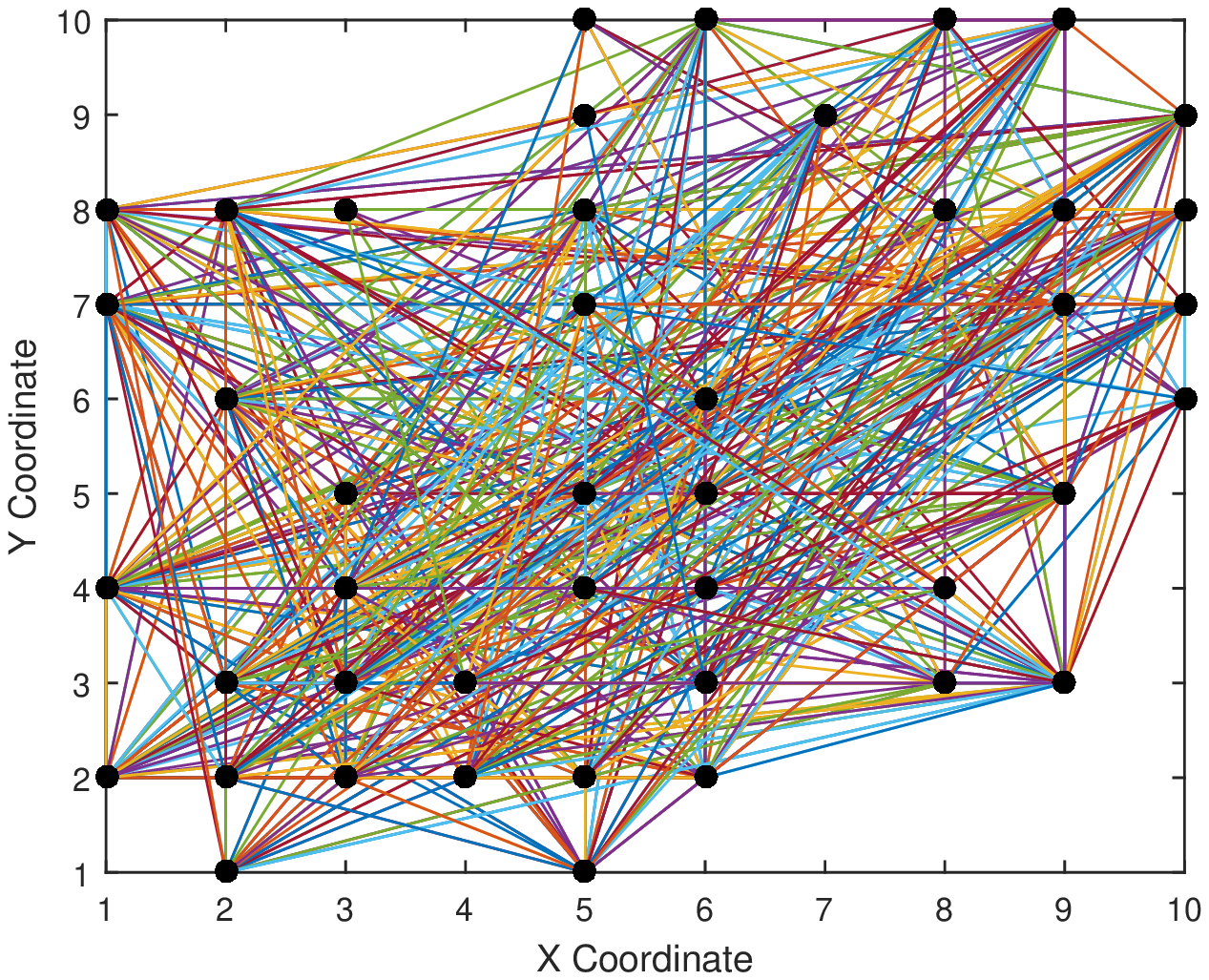}
}
\subfloat[$\{$ori 1,cue 2$\}$]{
\includegraphics[width=2.5in,height=2in]{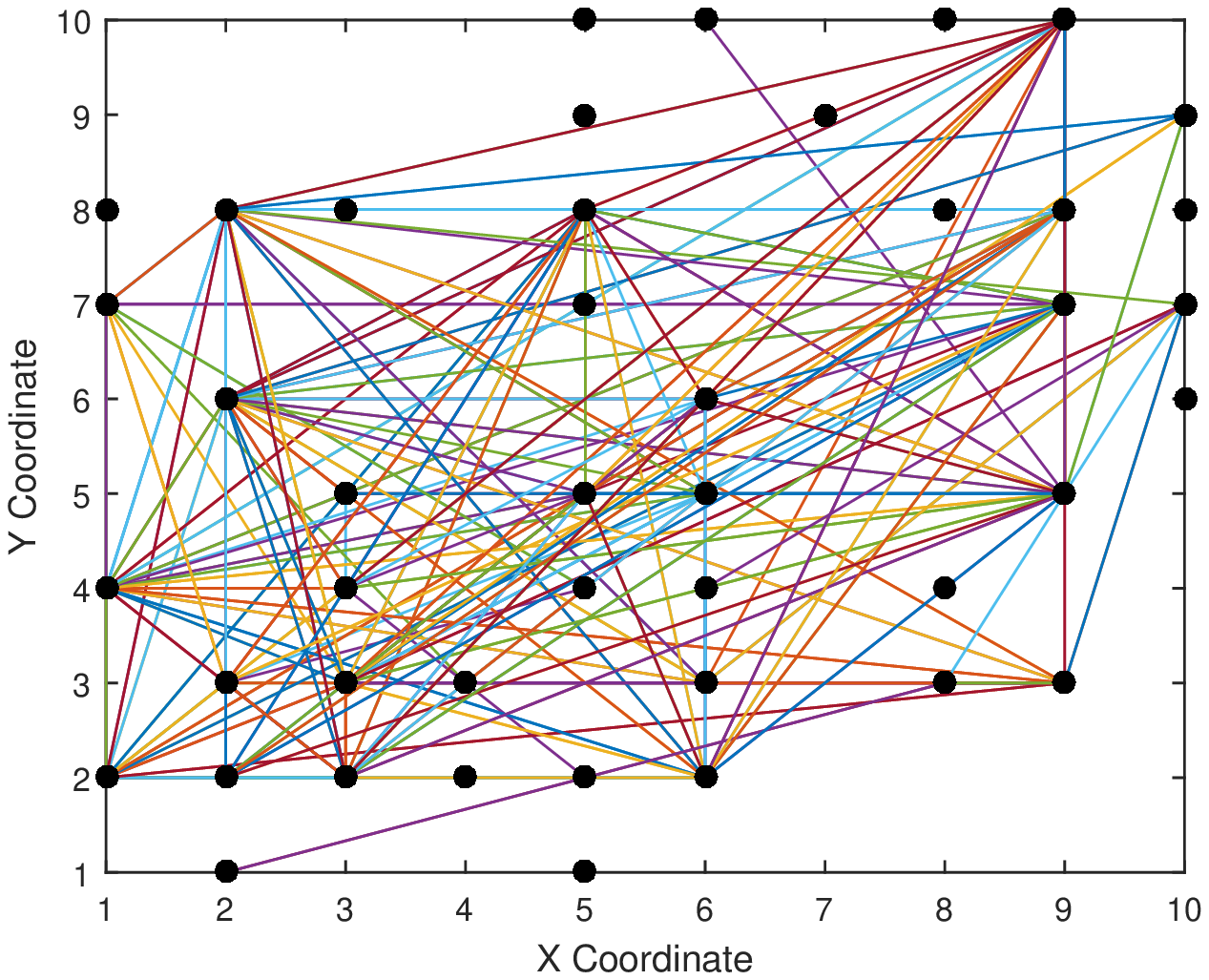}
}\\
\subfloat[$\{$ori 2,cue 1$\}$]{
\includegraphics[width=2.5in,height=2in]{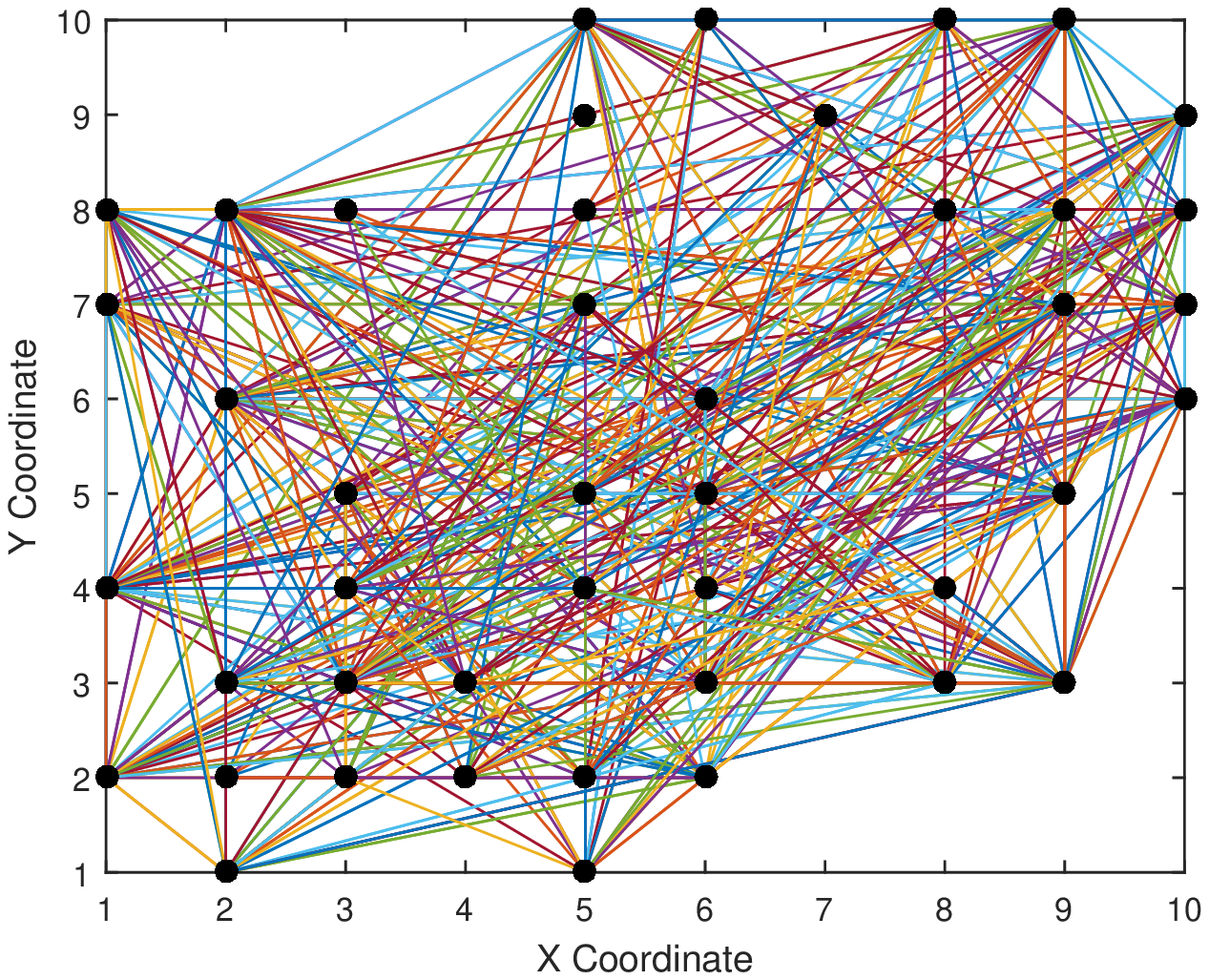}
}
\subfloat[$\{$ori 2,cue 2$\}$]{
\includegraphics[width=2.5in,height=2in]{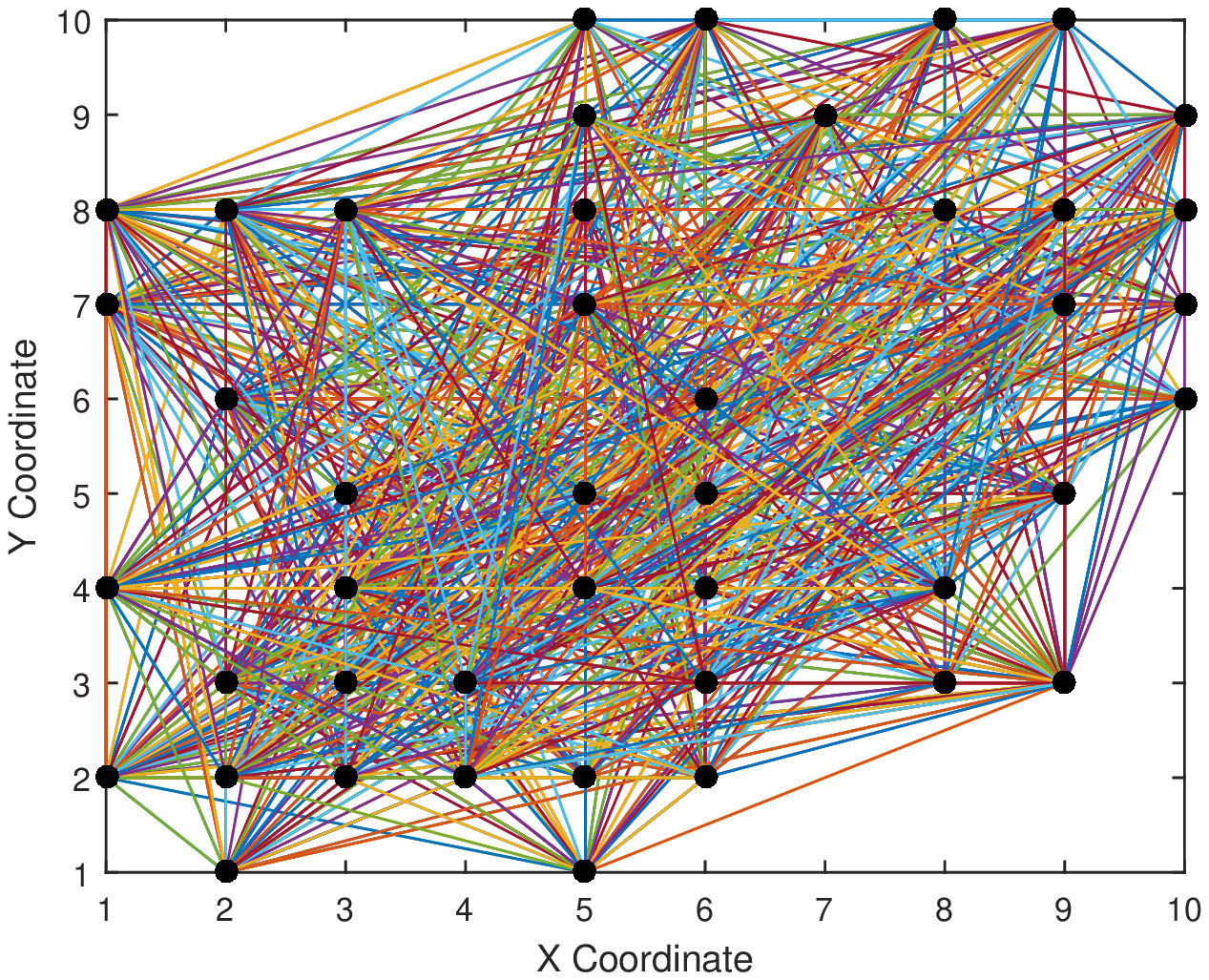}
}
\caption{The neuron network architecture detected by the SMC sampler. The four sub-figures depict the evolution of the structure at four different points in time, corresponding to four stimulus conditions $\{$ori 1, cue 1$\}$, $\{$ori 1, cue 2$\}$, $\{$ori 2, cue 1$\}$ and $\{$ori 2, cue 2$\}$, respectively.
The solid circles mark the positions of the involved neurons. Each line corresponds to a declared interaction between a pair of neurons.}\label{fig:network}
\end{figure}
%\begin{table}\centering\small
%\caption{Numbers of declared neuronal interactions across different stimulus conditions (condition 1=$\{$ori 1, cue 1$\}$, condition 2=$\{$ori 2, cue 1$\}$, condition 3=$\{$ori 1, cue 2$\}$, condition 4=$\{$ori 2, cue 2$\}$). The $(i,j)$th number in this table is the number of pairs of interactive neurons that are detected under both stimulus conditions indexed by $i$ and $j$}
%\begin{tabular}{c|c|c|c|c}
%\hline
%stimulus condition & 1 & 2 & 3 & 4 \\\hline
%1 & 899 & 553 & 282 & 885 \\\hline
%2 & 553 & 723 & 209 & 669 \\\hline
%3 & 282 & 209 & 282 & 281 \\\hline
%4 & 885 & 669 & 281 & 1722 \\\hline
%\end{tabular}
%\label{Table:num_of_detects}
%\end{table}

The neuron network architecture in terms of detected neuronal interactions is inferred from data and visually displayed in Fig.\ref{fig:network}.
%and from Table \ref{Table:num_of_detects}, we can check how many pairs of neurons change status (null vs. alternative) across stimulus conditions
%Corresponding to Fig.\ref{fig:network}.
Under four different stimulus conditions, corresponding to sub-figures (a), (b), (c), (d) of Fig.\ref{fig:network}, there are respectively 899, 723, 282 and 1722 pairwise neuronal interactions, which are detected by the SMC sampler algorithm. So it is confirmed that the neuron network structure varies long with the change in the stimulus condition.
Specifically, under the first stimulus condition $\{$ori 1, cue 1$\}$, namely, when the monkey was cued to attend towards the receptive field and the stimulus orientation was 135 degrees, the number of detected neural interactions is 899. Then, when the orientation was changed to be 45 degrees, corresponding to stimulus condition $\{$ori 2, cue 1$\}$, the number of detected interactions reduces to 723. Next the monkey was cued to away from the receptive field and the stimulus orientation was changed to be 135 degrees, namely the stimulus condition $\{$ori 1, cue 2$\}$ was applied. As a result, the number of detected neural interactions reduces to 282. Lastly, the stimulus orientation was changed to be 45 degrees, corresponding to stimulus condition $\{$ori 2, cue 2$\}$, and then the number of detected neural interactions increases rapidly from 282 to 1722. As is shown, the proposed algorithm allows the analyst to inspect the evolution of the neuron interaction network structure in real time, and the correlated pattern between the stimulus condition and the network structure can be analyzed in time.
\begin{figure}%[t]
\subfloat[$\{$ori 1, cue 1$\}$]{
\includegraphics[width=2in,height=1.5in]{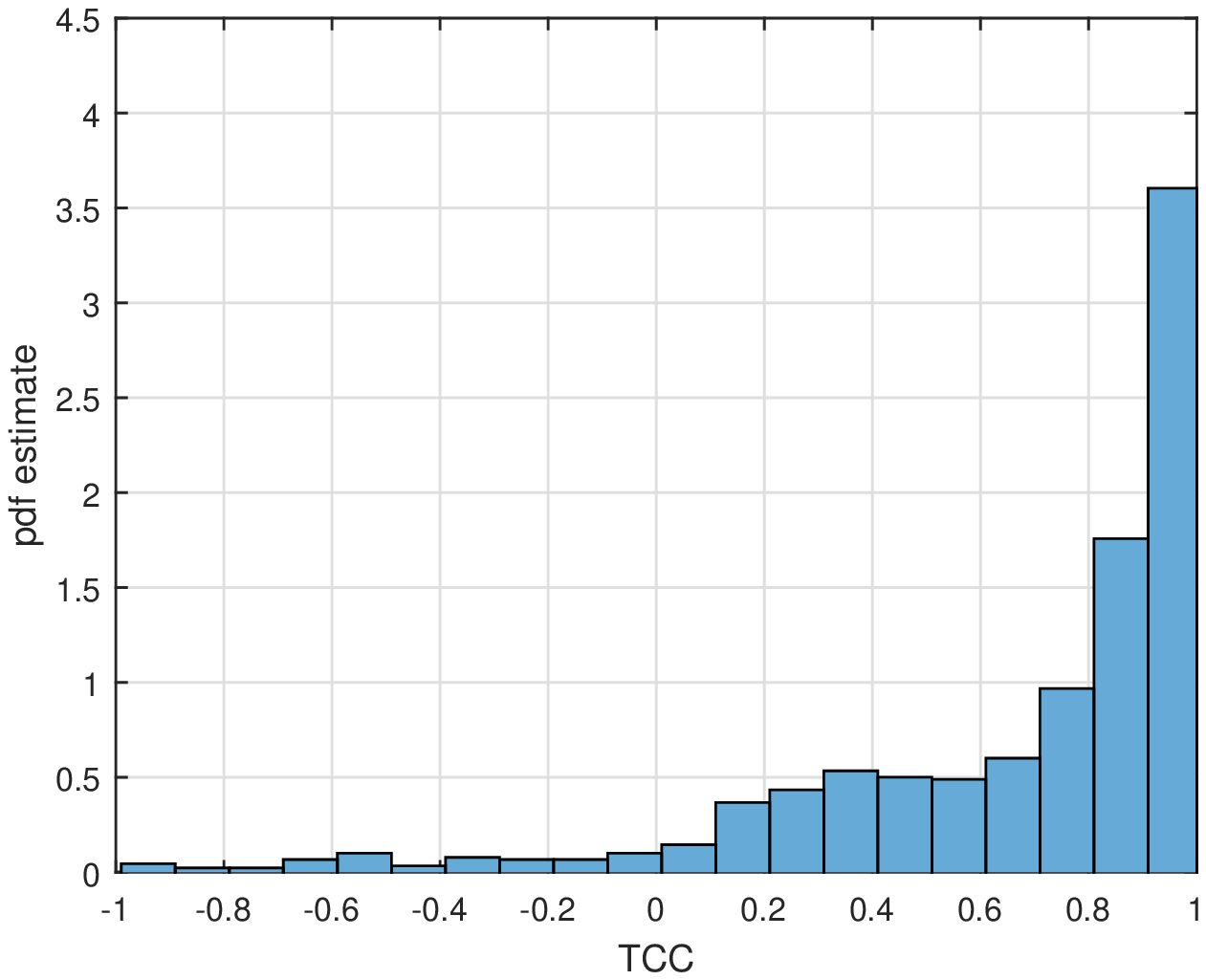}
}
\subfloat[$\{$ori 1, cue 1$\}$]{
\includegraphics[width=2in,height=1.5in]{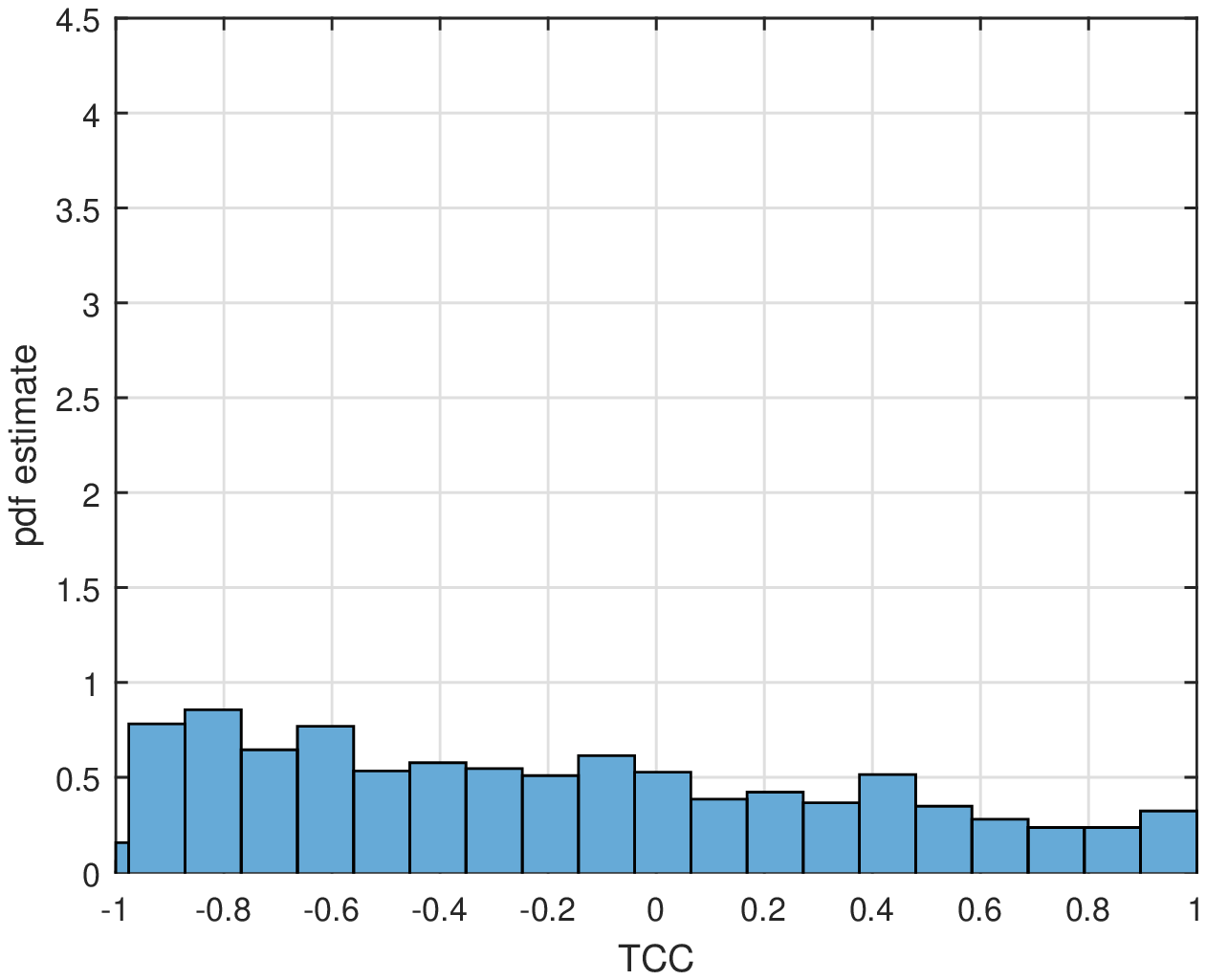}
}\\
\subfloat[$\{$ori 2, cue 1$\}$]{
\includegraphics[width=2in,height=1.5in]{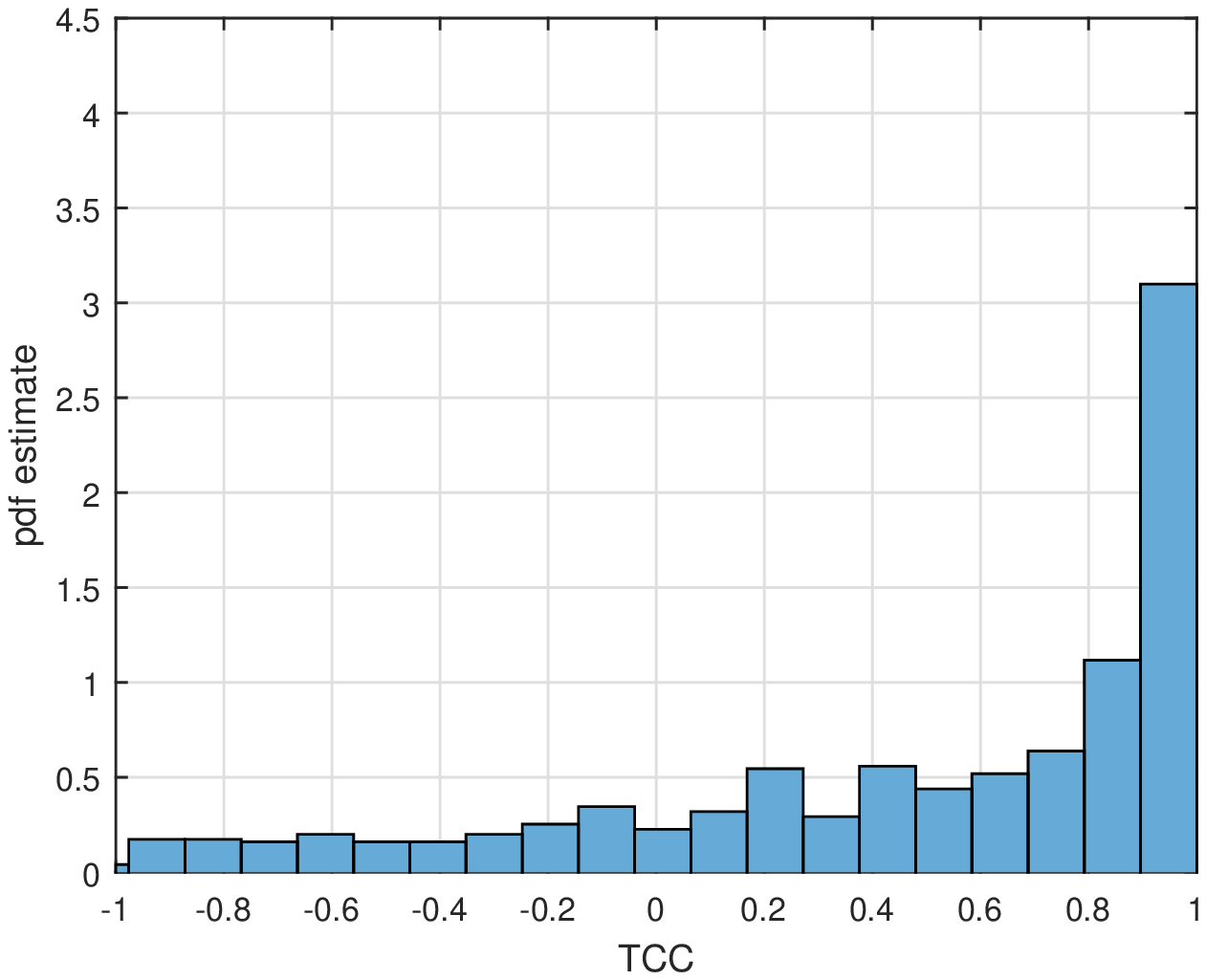}
}
\subfloat[$\{$ori 2, cue 1$\}$]{
\includegraphics[width=2in,height=1.5in]{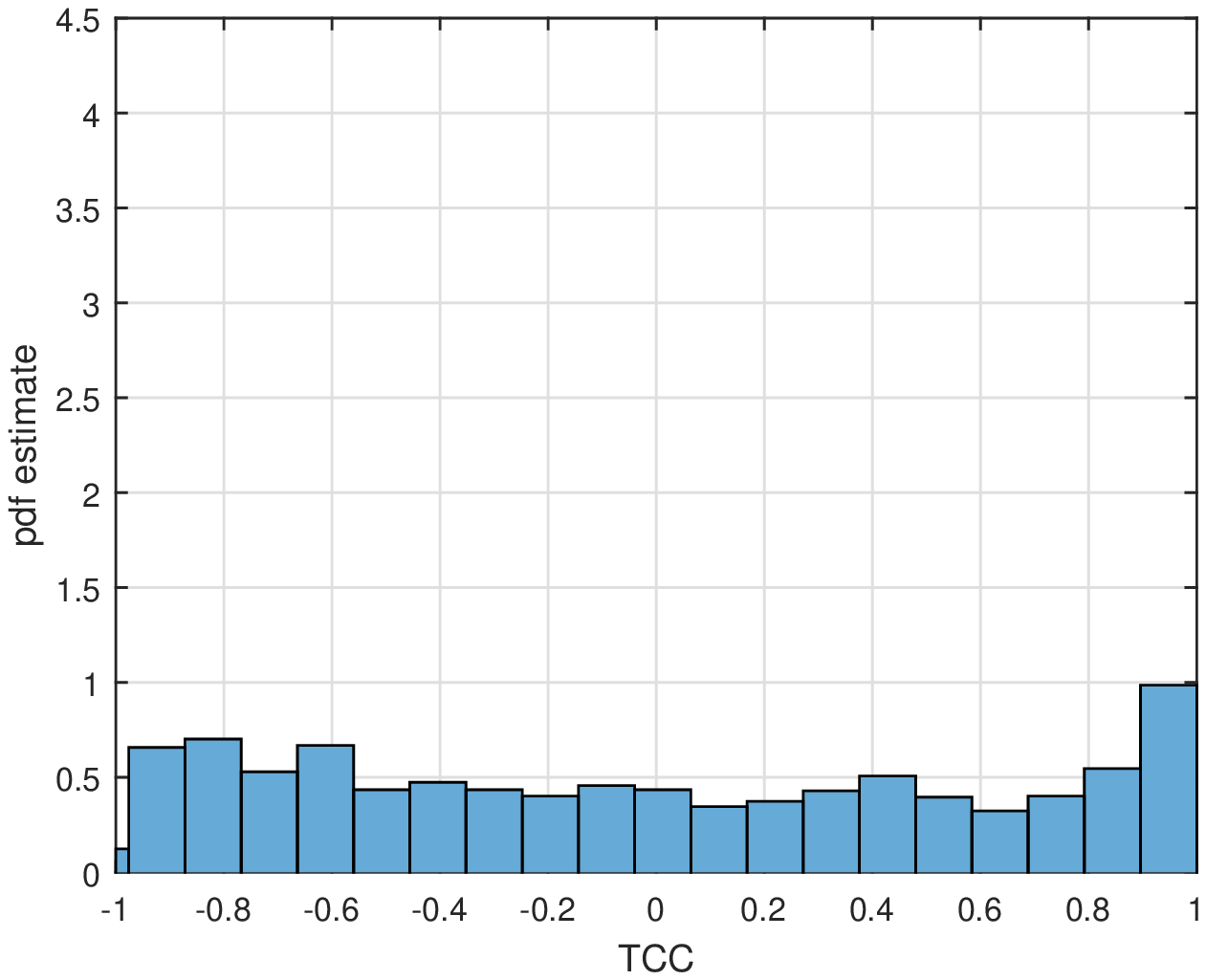}
}\\
\subfloat[$\{$ori 1, cue 2$\}$]{
\includegraphics[width=2in,height=1.5in]{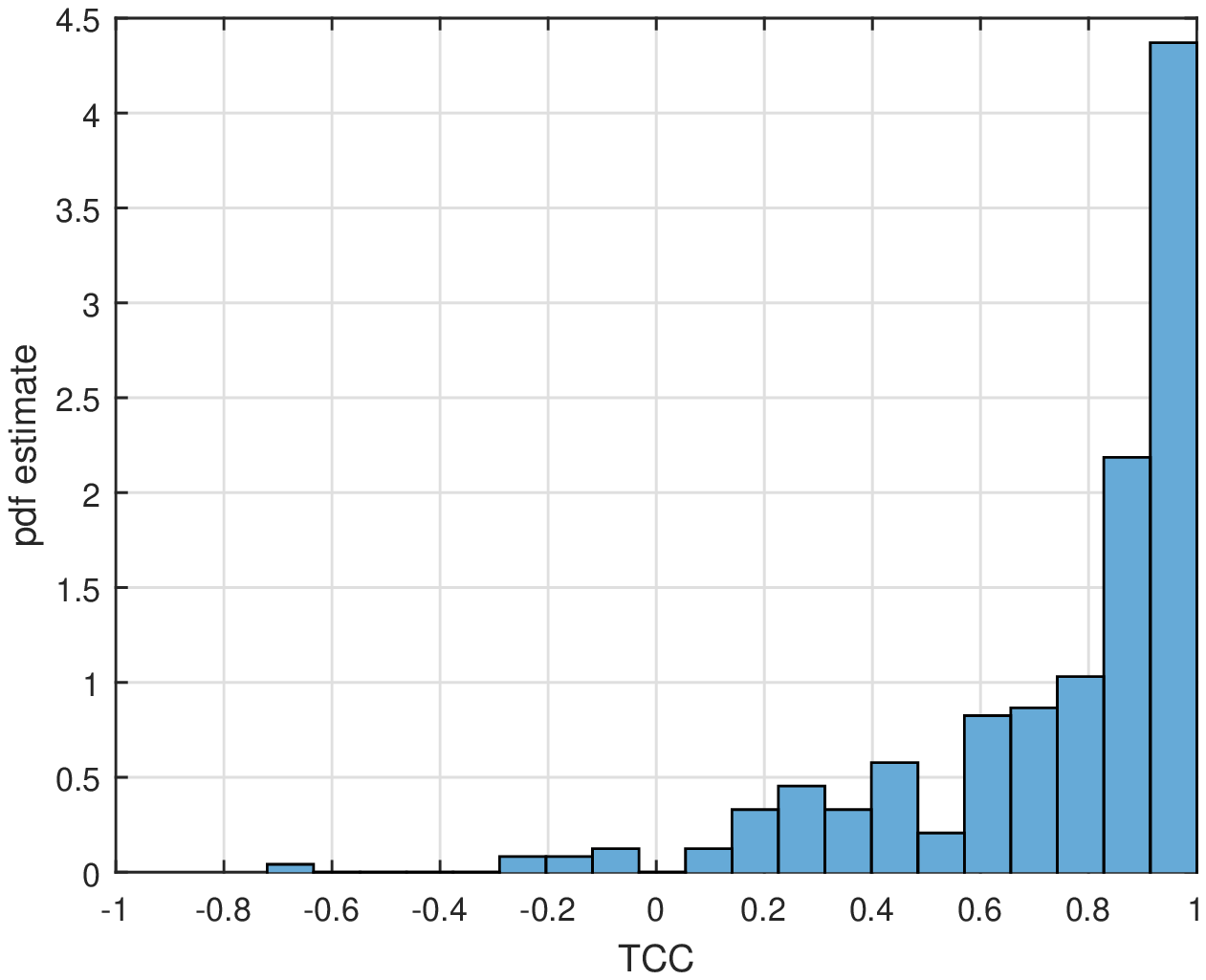}
}
\subfloat[$\{$ori 1, cue 2$\}$]{
\includegraphics[width=2in,height=1.5in]{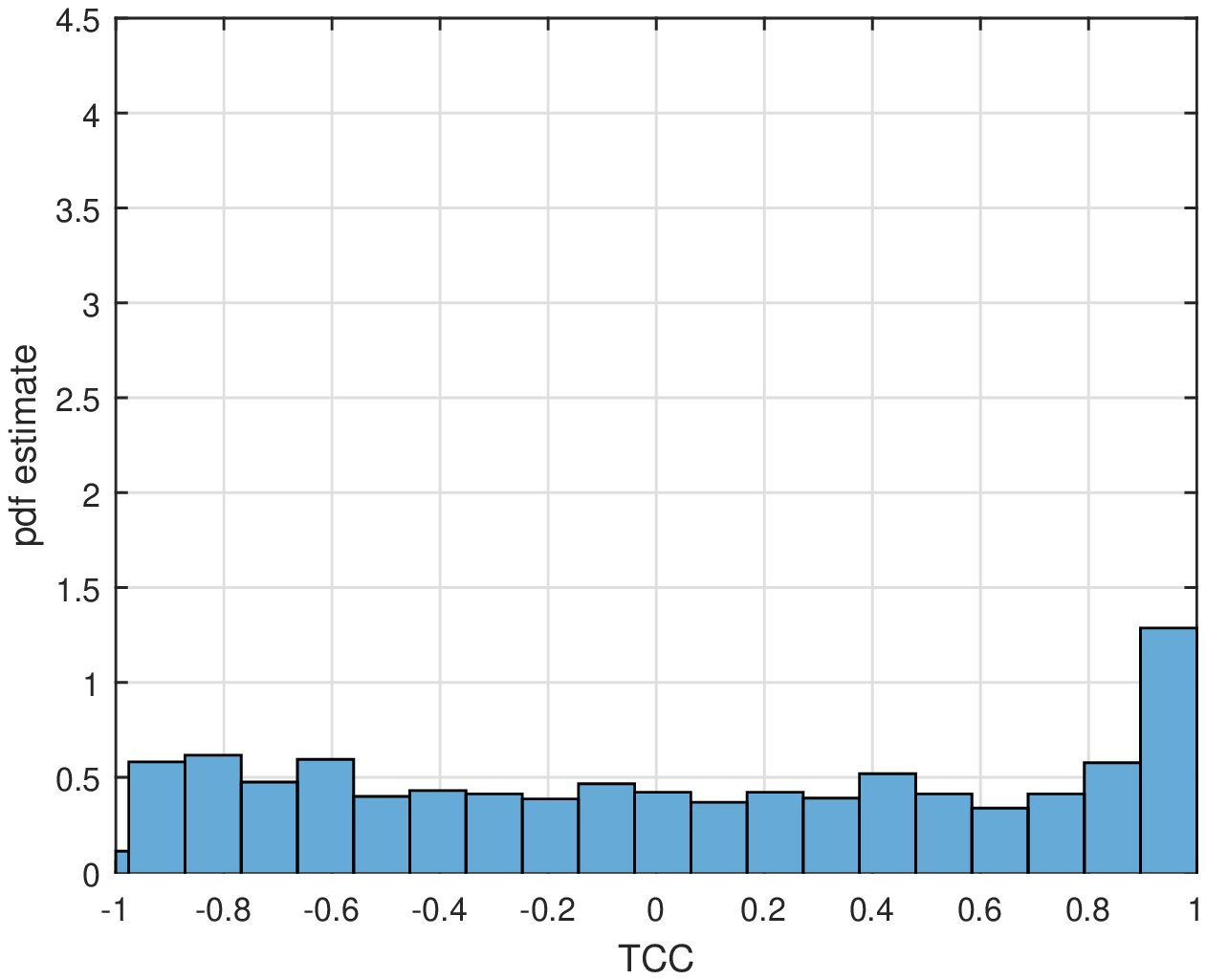}
}\\
\subfloat[$\{$ori 2, cue 2$\}$]{
\includegraphics[width=2in,height=1.5in]{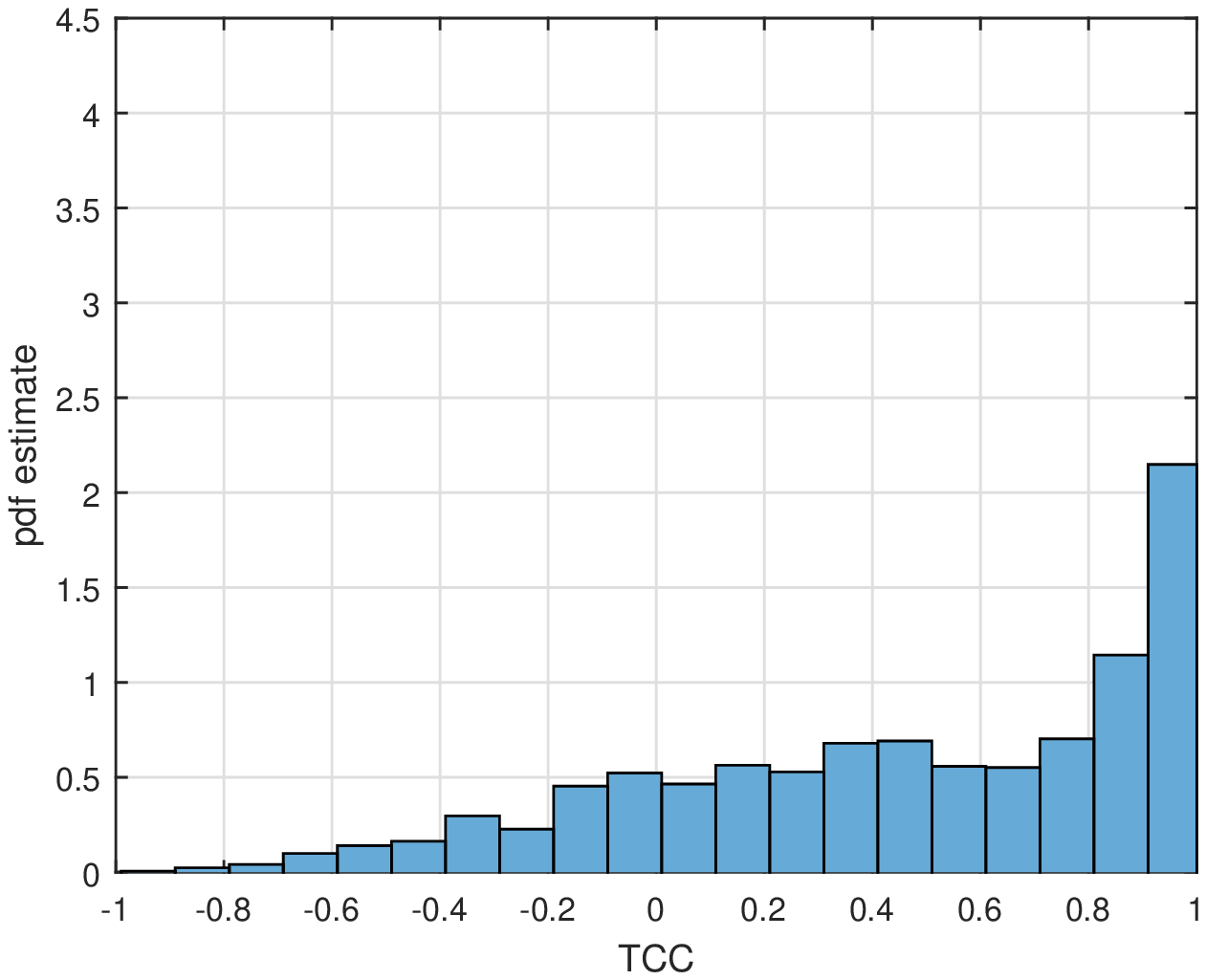}
}
\subfloat[$\{$ori 2, cue 2$\}$]{
\includegraphics[width=2in,height=1.5in]{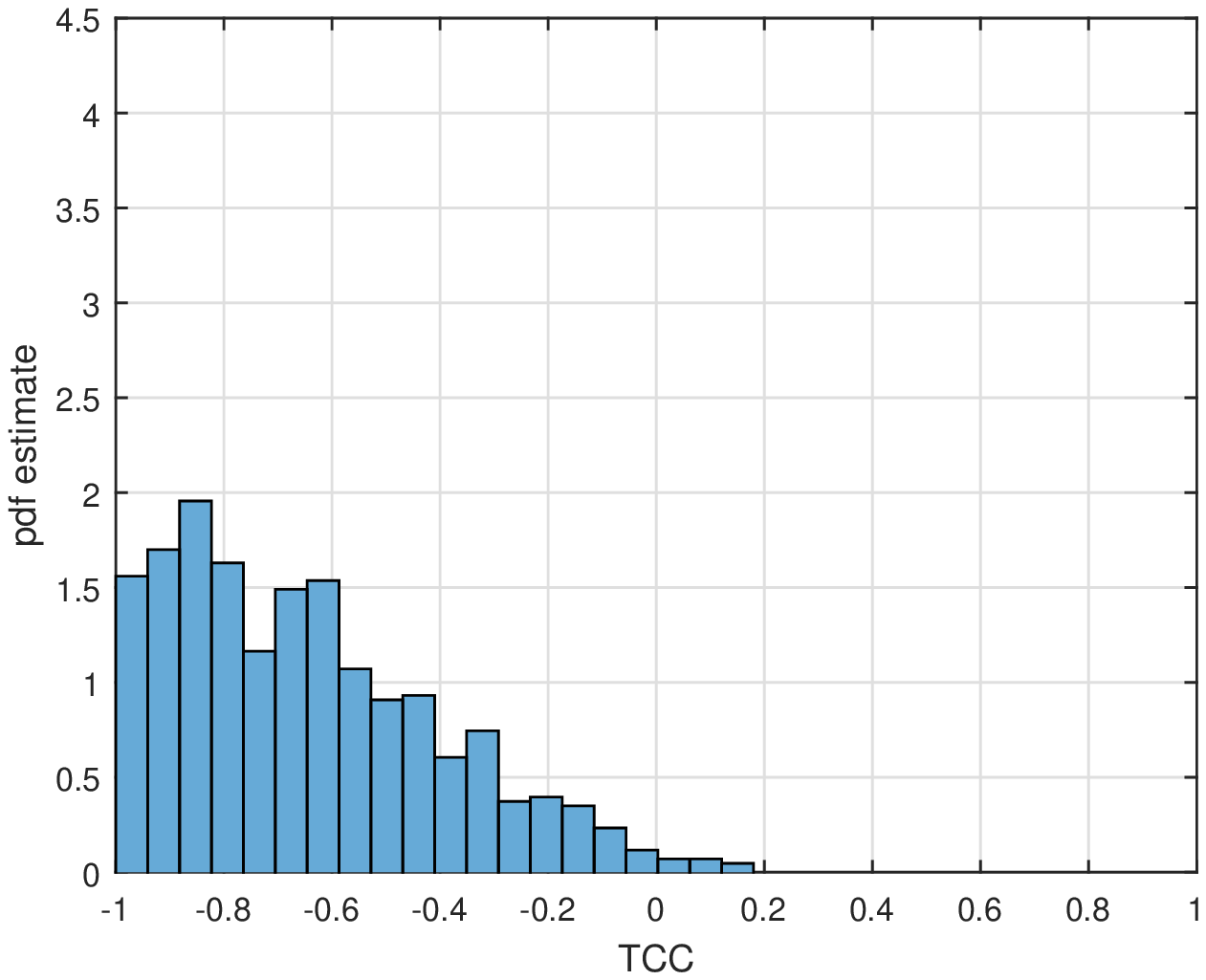}
}
\caption{The histograms of the turning curve correlation (TCC). The four rows of the sub-figures correspond to stimulus conditions $\{$ori 1, cue 1$\}$, $\{$ori 2, cue 1$\}$, $\{$ori 1, cue 2$\}$ and $\{$ori 2, cue 2$\}$, respectively. The two columns correspond to the declared alternative hypotheses and the declared null hypotheses, respectively.}\label{fig:hist_x2}
\end{figure}

Finally, we plot the histograms of the covariate TCC associated with different declared hypotheses, in Fig. \ref{fig:hist_x2}. We see that there is a clear correlation between the TCC and the values of the declared hypotheses, while the correlation structure varies along with the change in the stimulus condition. Specifically, under every stimulus condition, the distribution of the TCCs, associated with the declared alternative hypotheses, shrinks to the right, that is, the direction of 1; while the degrees of shrinkage associated with stimulus conditions with `ori 1' are slightly greater than those with `ori 2'.
Under all stimulus conditions except $\{$ori 2, cue 2$\}$, the distribution of the TCCs, associated with the declared null hypotheses, almost evenly covers the entire parameter space, while under stimulus condition $\{$ori 2,cue 2$\}$, this distribution shrinks to the left, that is, the direction of minus 1.

Traditional approaches have largely focused on a single scale to study the neuron network architecture. Although being illustrative, the above results indicate a potential of the proposed algorithm to expand beyond traditional approaches to capture the neuron network architecture on more scales, which include not only the time and space scales, but also the functional scale determined by the covariates.
\section{Conclusions}
We designed and implemented an SMC scheme for large scale multiple
testing in the context of BFDR, which is based on a hierarchical
parametric model that takes account of covariate information. We
showed that the performance of our algorithm can be as good as
the existing MCMC procedure while employing only a single pass
through the data, and that, after parallelization, it greatly reduces
computation time. We also illustrated the way this method could
be used with neural recordings to help identify network structure over time.
In other testing settings the particle rejuvenation step would need
to be re-designed, but the SMC scheme would still be applicable
and could easily be modified. The sequential scheme proposed here
thus opens the door to efficient computation in many large scale
multiple testing problems.
\section*{Acknowledgment}
%The authors would also like to thank the anonymous referees for their valuable comments and helpful suggestions.
This research has been supported by National NSFC Grant 61571238 (B. Liu), NIMH Grant RO1 064537 (R. E. Kass), NIH Grants 5R90DA023426-10 (G. Vinci), K99EY025768 (A. C. Snyder) and R01EY022928 (M. A. Smith).
\bibliographystyle{IEEEtran}
\bibliography{mybibfile}

\begin{thebibliography}{10}
\providecommand{\url}[1]{#1}
\csname url@samestyle\endcsname
\providecommand{\newblock}{\relax}
\providecommand{\bibinfo}[2]{#2}
\providecommand{\BIBentrySTDinterwordspacing}{\spaceskip=0pt\relax}
\providecommand{\BIBentryALTinterwordstretchfactor}{4}
\providecommand{\BIBentryALTinterwordspacing}{\spaceskip=\fontdimen2\font plus
\BIBentryALTinterwordstretchfactor\fontdimen3\font minus
  \fontdimen4\font\relax}
\providecommand{\BIBforeignlanguage}[2]{{%
\expandafter\ifx\csname l@#1\endcsname\relax
\typeout{** WARNING: IEEEtran.bst: No hyphenation pattern has been}%
\typeout{** loaded for the language `#1'. Using the pattern for}%
\typeout{** the default language instead.}%
\else
\language=\csname l@#1\endcsname
\fi
#2}}
\providecommand{\BIBdecl}{\relax}
\BIBdecl

\bibitem{bolton2003iterative}
R.~J. Bolton and N.~M. Adams, ``An iterative hypothesis-testing strategy for
  pattern discovery,'' in \emph{Proceedings of the ninth ACM SIGKDD
  international conference on Knowledge discovery and data mining}.\hskip 1em
  plus 0.5em minus 0.4em\relax ACM, 2003, pp. 49--58.

\bibitem{liu2015supporting}
G.~Liu, H.~Zhang, M.~Feng, L.~Wong, and S.-K. Ng, ``Supporting exploratory
  hypothesis testing and analysis,'' \emph{ACM Transactions on Knowledge
  Discovery from Data (TKDD)}, vol.~9, no.~4, p.~31, 2015.

\bibitem{webb2016multiple}
G.~I. Webb and F.~Petitjean, ``A multiple test correction for streams and
  cascades of statistical hypothesis tests,'' in \emph{Proc. of the 22nd ACM
  SIGKDD Int'l Conf. on Knowledge Discovery and Data Mining}.\hskip 1em plus
  0.5em minus 0.4em\relax ACM, 2016, pp. 1255--1264.

\bibitem{liu2006statistical}
C.~Liu, L.~Fei, X.~Yan, J.~Han, and S.~P. Midkiff, ``Statistical debugging: A
  hypothesis testing-based approach,'' \emph{IEEE Transactions on Software
  Engineering}, vol.~32, no.~10, pp. 831--848, 2006.

\bibitem{ge2003resampling}
Y.~Ge, S.~Dudoit, and T.~P. Speed, ``Resampling-based multiple testing for
  microarray data analysis,'' \emph{Test}, vol.~12, no.~1, pp. 1--77, 2003.

\bibitem{ignatiadis2016data}
N.~Ignatiadis, B.~Klaus, J.~B. Zaugg, and W.~Huber, ``Data-driven hypothesis
  weighting increases detection power in genome-scale multiple testing,''
  \emph{Nature methods}, vol.~13, no.~7, pp. 577--580, 2016.

\bibitem{durante2016bayesian}
D.~Durante and D.~B. Dunson, ``Bayesian inference and testing of group
  differences in brain networks,'' \emph{Bayesian Analysis}, 2016.

\bibitem{kohavi2009controlled}
R.~Kohavi, R.~Longbotham, D.~Sommerfield, and R.~M. Henne, ``Controlled
  experiments on the web: survey and practical guide,'' \emph{Data mining and
  knowledge discovery}, vol.~18, no.~1, pp. 140--181, 2009.

\bibitem{armitage2008statistical}
P.~Armitage, G.~Berry, and J.~N.~S. Matthews, \emph{Statistical methods in
  medical research}.\hskip 1em plus 0.5em minus 0.4em\relax John Wiley \& Sons,
  2008.

\bibitem{bender2001adjusting}
R.~Bender and S.~Lange, ``Adjusting for multiple testing—when and how?''
  \emph{Journal of clinical epidemiology}, vol.~54, no.~4, pp. 343--349, 2001.

\bibitem{benjamini1995controlling}
Y.~Benjamini and Y.~Hochberg, ``Controlling the false discovery rate: a
  practical and powerful approach to multiple testing,'' \emph{Journal of the
  royal statistical society. Series B (Methodological)}, pp. 289--300, 1995.

\bibitem{efron2012large}
B.~Efron, \emph{Large-scale inference: empirical Bayes methods for estimation,
  testing, and prediction}.\hskip 1em plus 0.5em minus 0.4em\relax Cambridge
  University Press, 2012, vol.~1.

\bibitem{muller2006fdr}
P.~Muller, G.~Parmigiani, and K.~Rice, ``Fdr and bayesian multiple comparisons
  rules,'' \emph{Johns Hopkins University, Dept. of Biostatistics Working
  Papers}, 2006.

\bibitem{scott2015false}
J.~G. Scott, R.~C. Kelly, M.~A. Smith, P.~Zhou, and R.~E. Kass, ``False
  discovery rate regression: an application to neural synchrony detection in
  primary visual cortex,'' \emph{Journal of the American Statistical
  Association}, vol. 110, no. 510, pp. 459--471, 2015.

\bibitem{smith2013spatial}
M.~A. Smith and M.~A. Sommer, ``Spatial and temporal scales of neuronal
  correlation in visual area v4,'' \emph{Journal of Neuroscience}, vol.~33,
  no.~12, pp. 5422--5432, 2013.

\bibitem{saalmann2011cognitive}
Y.~B. Saalmann and S.~Kastner, ``Cognitive and perceptual functions of the
  visual thalamus,'' \emph{Neuron}, vol.~71, no.~2, pp. 209--223, 2011.

\bibitem{smith2008spatial}
M.~A. Smith and A.~Kohn, ``Spatial and temporal scales of neuronal correlation
  in primary visual cortex,'' \emph{Journal of Neuroscience}, vol.~28, no.~48,
  pp. 12\,591--12\,603, 2008.

\bibitem{averbeck2006neural}
B.~B. Averbeck, P.~E. Latham, and A.~Pouget, ``Neural correlations, population
  coding and computation,'' \emph{Nature reviews neuroscience}, vol.~7, no.~5,
  pp. 358--366, 2006.

\bibitem{cohen2011measuring}
M.~R. Cohen and A.~Kohn, ``Measuring and interpreting neuronal correlations,''
  \emph{Nature neuroscience}, vol.~14, no.~7, pp. 811--819, 2011.

\bibitem{doiron2016mechanics}
B.~Doiron, A.~Litwin-Kumar, R.~Rosenbaum, G.~K. Ocker, and K.~Josi{\'c}, ``The
  mechanics of state-dependent neural correlations,'' \emph{Nature
  neuroscience}, vol.~19, no.~3, pp. 383--393, 2016.

\bibitem{yatsenko2015improved}
D.~Yatsenko, K.~Josi{\'c}, A.~S. Ecker, E.~Froudarakis, R.~J. Cotton, and A.~S.
  Tolias, ``Improved estimation and interpretation of correlations in neural
  circuits,'' \emph{PLoS Comput Biol}, vol.~11, no.~3, p. e1004083, 2015.

\bibitem{chopin2002sequential}
N.~Chopin, ``A sequential particle filter method for static models,''
  \emph{Biometrika}, vol.~89, no.~3, pp. 539--552, 2002.

\bibitem{del2006sequential}
P.~Del~Moral, A.~Doucet, and A.~Jasra, ``Sequential monte carlo samplers,''
  \emph{Journal of the Royal Statistical Society: Series B (Statistical
  Methodology)}, vol.~68, no.~3, pp. 411--436, 2006.

\bibitem{arulampalam2002tutorial}
M.~S. Arulampalam, S.~Maskell, N.~Gordon, and T.~Clapp, ``A tutorial on
  particle filters for online nonlinear/non-gaussian bayesian tracking,''
  \emph{IEEE Transactions on signal processing}, vol.~50, no.~2, pp. 174--188,
  2002.

\bibitem{gilks2001following}
W.~R. Gilks and C.~Berzuini, ``Following a moving target—monte carlo
  inference for dynamic bayesian models,'' \emph{Journal of the Royal
  Statistical Society: Series B (Statistical Methodology)}, vol.~63, no.~1, pp.
  127--146, 2001.

\bibitem{Li2015Resampling}
T.~Li, M.~Bolic, and P.~M. Djuric, ``Resampling methods for particle filtering:
  classification, implementation, and strategies,'' \emph{IEEE Signal
  Processing Magazine}, vol.~32, no.~3, pp. 70--86, 2015.

\bibitem{Hol2006on}
J.~D. Hol, T.~B. Schon, and G.~F., ``On resampling algorithms for particle
  filters,'' in \emph{Proc. of the IEEE Nonlinear Statistical Signal Processing
  Workshop (NSSPW)}.\hskip 1em plus 0.5em minus 0.4em\relax IEEE, 2006, pp.
  79--82.

\bibitem{crisan2002survey}
D.~Crisan and A.~Doucet, ``A survey of convergence results on particle
  filtering methods for practitioners,'' \emph{IEEE Transactions on signal
  processing}, vol.~50, no.~3, pp. 736--746, 2002.

\bibitem{bengtsson2008curse}
T.~Bengtsson, P.~Bickel, B.~Li \emph{et~al.}, ``Curse-of-dimensionality
  revisited: Collapse of the particle filter in very large scale systems,'' in
  \emph{Probability and statistics: Essays in honor of David A.
  Freedman}.\hskip 1em plus 0.5em minus 0.4em\relax Institute of Mathematical
  Statistics, 2008, pp. 316--334.

\bibitem{liu1998sequential}
J.~S. Liu and R.~Chen, ``Sequential monte carlo methods for dynamic systems,''
  \emph{Journal of the American statistical association}, vol.~93, no. 443, pp.
  1032--1044, 1998.

\bibitem{gordon1993novel}
N.~J. Gordon, D.~J. Salmond, and A.~F. Smith, ``Novel approach to
  nonlinear/non-gaussian bayesian state estimation,'' in \emph{IEE Proceedings
  F (Radar and Signal Processing)}, vol. 140, no.~2.\hskip 1em plus 0.5em minus
  0.4em\relax IET, 1993, pp. 107--113.

\bibitem{carpenter1999improved}
J.~Carpenter, P.~Clifford, and P.~Fearnhead, ``Improved particle filter for
  nonlinear problems,'' \emph{IEE Proceedings-Radar, Sonar and Navigation},
  vol. 146, no.~1, pp. 2--7, 1999.

\bibitem{stauffer1999adaptive}
C.~Stauffer and W.~E.~L. Grimson, ``Adaptive background mixture models for
  real-time tracking,'' in \emph{Computer Vision and Pattern Recognition, 1999.
  IEEE Computer Society Conference on.}, vol.~2.\hskip 1em plus 0.5em minus
  0.4em\relax IEEE, 1999, pp. 246--252.

\bibitem{balakrishnan2006one}
S.~Balakrishnan, D.~Madigan \emph{et~al.}, ``A one-pass sequential monte carlo
  method for bayesian analysis of massive datasets,'' \emph{Bayesian Analysis},
  vol.~1, no.~2, pp. 345--361, 2006.

\bibitem{stavropoulos2001improved}
P.~Stavropoulos and D.~Titterington, ``Improved particle filters and
  smoothing,'' in \emph{Sequential Monte Carlo Methods in Practice}.\hskip 1em
  plus 0.5em minus 0.4em\relax Springer, 2001, pp. 295--317.

\bibitem{silverman1986density}
B.~W. Silverman, \emph{Density estimation for statistics and data
  analysis}.\hskip 1em plus 0.5em minus 0.4em\relax CRC press, 1986, vol.~26.

\bibitem{albert2012combining}
I.~Albert, S.~Donnet, C.~Guihenneuc-Jouyaux, S.~Low-Choy, K.~Mengersen,
  J.~Rousseau \emph{et~al.}, ``Combining expert opinions in prior
  elicitation,'' \emph{Bayesian Analysis}, vol.~7, no.~3, pp. 503--532, 2012.

\bibitem{dey2012practical}
D.~D. Dey, P.~M{\"u}Iler, and D.~Sinha, \emph{Practical nonparametric and
  semiparametric Bayesian statistics}.\hskip 1em plus 0.5em minus 0.4em\relax
  Springer Science \& Business Media, 2012, vol. 133.

\bibitem{gelman2014bayesian}
A.~Gelman, J.~B. Carlin, H.~S. Stern, and D.~B. Rubin, \emph{Bayesian data
  analysis}.\hskip 1em plus 0.5em minus 0.4em\relax Chapman \& Hall/CRC Boca
  Raton, FL, USA, 2014, vol.~2.

\bibitem{efron2004large}
B.~Efron, ``Large-scale simultaneous hypothesis testing: the choice of a null
  hypothesis,'' \emph{Journal of the American Statistical Association},
  vol.~99, no. 465, pp. 96--104, 2004.

\bibitem{kass2014analysis}
R.~E. Kass, U.~T. Eden, and E.~N. Brown, \emph{Analysis of neural data}.\hskip
  1em plus 0.5em minus 0.4em\relax Springer, 2014, vol. 491.

\bibitem{vinci2016separating}
G.~Vinci, V.~Ventura, M.~A. Smith, and R.~E. Kass, ``Separating spike count
  correlation from firing rate correlation,'' \emph{Neural computation},
  vol.~28, pp. 849--881, 2016.

\bibitem{kong1994sequential}
A.~Kong, J.~S. Liu, and W.~H. Wong, ``Sequential imputations and bayesian
  missing data problems,'' \emph{Journal of the American statistical
  association}, vol.~89, no. 425, pp. 278--288, 1994.

\end{thebibliography}
\end{document}